\documentclass[%
 reprint,
 amsmath,amssymb,
 aps,soul,
pre,
]{revtex4-2}

\usepackage{graphicx}
\usepackage{dcolumn}
\usepackage{bm}
\usepackage{xcolor}

\renewcommand{\selectlanguage}[1]{}
\usepackage{lipsum}
\makeatletter
\newcommand*{\balancecolsandclearpage}{%
  \close@column@grid
  \cleardoublepage
  \twocolumngrid
}
\begin{document}

\preprint{APS/123-QED}

\title{Eastward Transients in the Dayside Ionosphere II: \\ A Parallel-plate Capacitor-Like Effect}

\author{Magnus F Ivarsen}
\altaffiliation[Also at ]{
The European Space Agency Centre for Earth Observation, Frascati, Italy}
\affiliation{Department of Physics and Engineering Physics, University of Saskatchewan, Saskatoon, Canada}%

\author{Yaqi Jin}
\affiliation{Department of Physics, University of Oslo, Oslo, Norway}

\author{Devin R Huyghebaert}
\altaffiliation[Also at ]{Department of Physics and Engineering Physics, University of Saskatchewan, Saskatoon, Canada}
\affiliation{Leibniz Institute of Atmospheric Physics, K\"{u}hlungsborn, Germany}%

\author{Jean-Pierre St-Maurice}
\altaffiliation[Also at ]{Department of Physics and Astronomy, University of Western Ontario, London, Ontario, Canada}
\author{Glenn C Hussey}
\author{Kathryn McWilliams}
\affiliation{Department of Physics and Engineering Physics, University of Saskatchewan, Saskatoon, Canada}

\author{Yukinaga Miyashita}
\altaffiliation[Also at ]{Department of Astronomy and Space Science, Korea University of Science and Technology, Daejeon, South-Korea}
\affiliation{Space Science Division, Korea Astronomy and Space Science Institute, Daejeon, South-Korea}%

\author{David Sibeck}
\affiliation{Goddard Space Flight Center, NASA, Greenbelt, USA}

\date{\today}

\begin{abstract}
During the 23 April 2023 geospace storm, we observed chorus wave-driven, energetic particle precipitation on closed magnetic field lines in the dayside magnetosphere. Simultaneously and in the ionosphere's bottom-side, we observed signatures of impact ionization and strong enhancements in the ionospheric electric field, via radar-detection of meter-scale turbulence, and with matching temporal characteristics as that of the magnetospheric observations. We detailed this in a companion paper. In the present article, we place those observations into context with the dayside ionosphere, and describe a remarkably similar event that took place during the May 2024 geospace superstorm. In both cases, fast, eastward-moving electric field structures were excited equatorward of the ionospheric cusp, on closed magnetic field-lines -- observations that challenge existing modes of explanation for electrodynamics in the cusp-region, where most such observations are interpreted in the context of poleward-moving auroral forms. Instead, primarily eastward-moving electric field structures were associated with turbulent Hall currents that are perhaps characteristically excited during geospace storms by wave-particle interactions near magnetospheric equator or by proton precipitation characteristics in the cusp, forming a `parallel-plate capacitor-like effect'. We propose that transient eastward electrodynamic bursts in the dayside ionosphere might be a common, albeit previously unresolved, feature of geomagnetic storms.
\end{abstract}

\maketitle



\section{\label{sec:intro}Introduction}

The solar wind, a stream of plasma at various densities, pushes against Earth's magnetosphere, driving large-scale electrical currents and plasma convection in Earth's ionosphere \cite{dungeyInterplanetaryMagneticField1961}. Large energy transfers are brought about through magnetic reconnection, the interface between the terrestrial and solar magnetic fields, a process that transfers magnetic flux from the dayside to the nightside  \cite{cowleyTUTORIALMagnetosphereIonosphereInteractions2000}. On the Sun-facing side of the magnetosphere, closed field-lines open up to the solar wind, forming the magnetospheric cleft, or \emph{cusp} region \cite{saundersPolarCuspIonosphere1989}, which plays host to plasma outflow and particle precipitation \cite{shepherdDaysideCleftAurora1979,ogawaSimultaneousEISCATSvalbard2003}.

\begin{figure}
    \centering
    \includegraphics[width=0.5\textwidth]{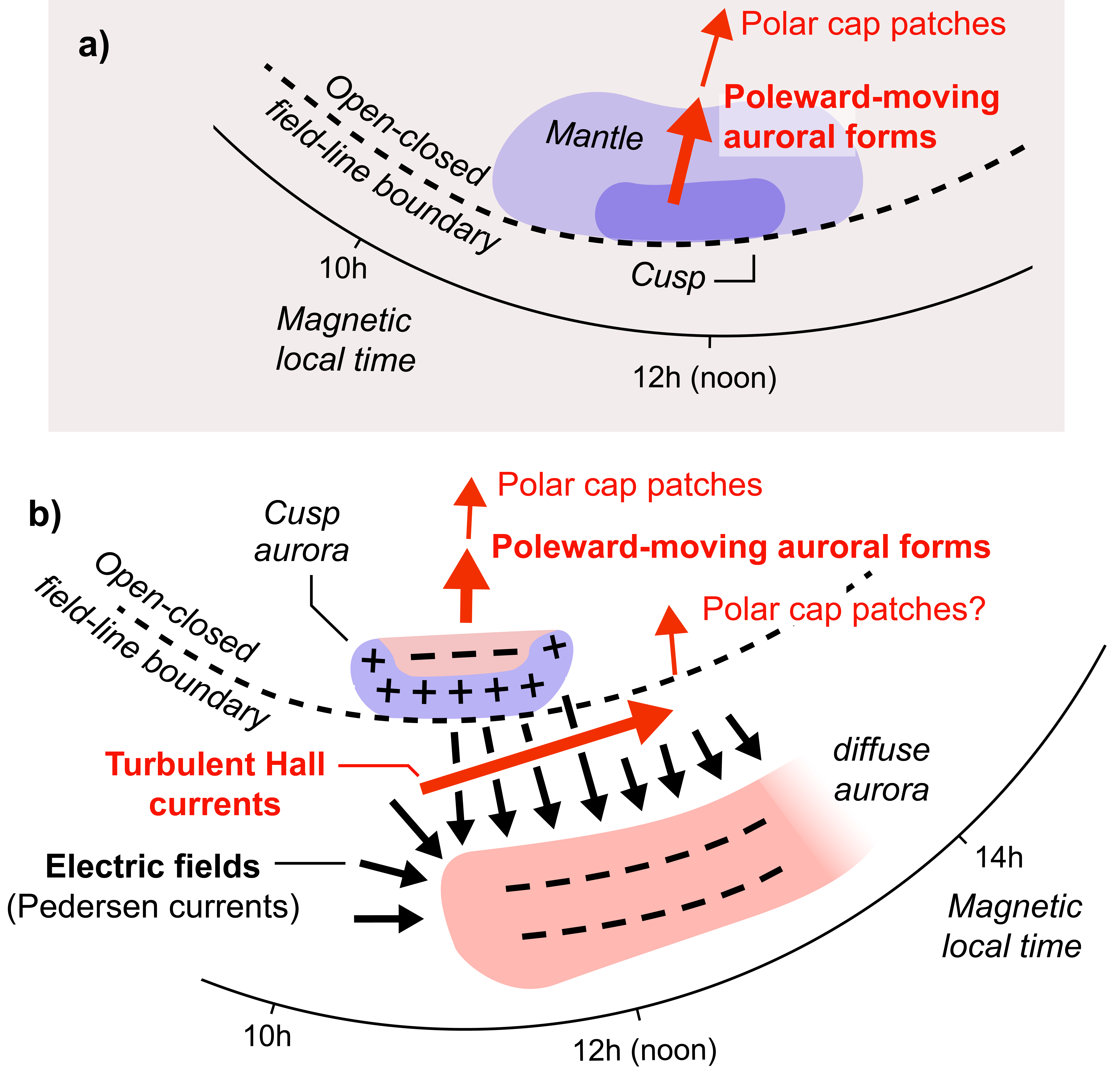}
    \caption{
    \textbf{Panel a)}: Schematic drawing of the cusp, its extended mantle, and the motion of traditional poleward-moving aurora forms, as well as their expected seeding of polar cap patches. 
    \textbf{Panel b)}: Schematic drawing of the greater cusp-region, showing a mix of low-energy electrons (pink shaded area, minus signs) and protons (light blue shaded area, plus signs), with the most intense proton aurora on its equatorward edge. A region of high-energy diffuse aurora lies somewhere to the southeast of the cusp, and the two regions are separated by the open-closed field-line boundary. An strong equatorward electric field forms between the regions, driving turbulent electrojets, currents whose laminar form is broken up and developing the condition. The action of the turbulent Hall drifts (red arrow) may push islands of structured ionization into the polar cap, if aided by dayside reconnection.
    }
    \label{fig:cartoon}
\end{figure}

Poleward-moving auroral forms are frequently sighted in the ionospheric cusp and they are manifestations of dayside magnetic reconnection. They consist mostly of red auroral arcs that drift poleward along with the reconnected field-lines from the cusp and into the polar cap \emph{in pulses} \cite{frey_dayside_2019,southwoodWhatAreFlux1988,oksavikHighresolutionObservationsSmallscale2004}, as is illustrated in Figure~\ref{fig:cartoon}a). The pulsations reflect the temporal evolution in flux transfer events, meaning that a study of the dynamics in one system is applicable to the other \cite{faselDaysidePolewardMoving1995,lester_coherent-scatter_1998}. For this reason, the systematic study of poleward-moving auroral forms have yielded crucial physical insight into the magnetosphere. Relevant for the study at hand, such aurorae are invariably accompanied by fast plasma flow channels in the topside ionosphere \cite{sandholtPolewardMovingAuroral2007,oksavikScintillationLossSignal2015}, with typical reported speeds ranging from 500~m/s to 2~km/s \cite{milanConvectionAuroralResponse2000}.

There are however observations of poleward-moving forms that move faster than the magnetospheric convection would imply \cite{mcwilliams_two-dimensional_2000}, in counter-intuitive directions \cite{sandholtAuroralElectrodynamicsCusp1991}, and some that occur on \textit{closed} magnetic field-lines \cite{kozlovsky_motion_2002}, observations that challenge the accepted notion of magnetic flux being peeled off from the magnetopause into the polar cap \cite{wu_simultaneous_2020}. In an attempt at reconciling the field, Ref.~\cite{lyatsky_central_1997} proposed that hot electrons from a dayside plasma sheet \cite{lyatsky_penetration_2016} may precipitate in numbers sufficient to produce a perturbation electric field (aligned with the Pedersen currents that close the field-aligned currents), thereby powering fast drifts, a situation that we illustrate in Figure~\ref{fig:cartoon}b).

Indeed, high-energy diffuse aurorae are a staple of the dayside ionosphere \cite{spasojevic_drivers_2010,nishimura_structures_2013,ni_chorus_2014}, where the mechanisms responsible originate with plasma waves near the magnetospheric equator; these interact with orbiting electrons, causing pitch-angle scattering of the those electrons into the loss cone, and subsequent precipitation into Earth's atmosphere \cite{ni_resonant_2008,ni_resonant_2011}. Having high kinetic energy, dayside diffuse aurorae will ionize the E-region, the bottomside ionosphere \cite{fangParameterizationMonoenergeticElectron2010}, at altitudes even lower than the extant extreme ultra-violet radiation from Sun would imply \cite{prolssAbsorptionDissipationSolar2004}. Thereby, diffuse aurorae drive field-aligned currents, an action whose reaction comes in the form of local, strong, perpendicular electric fields, conforming to expectations from the conservation of current vorticity, fields that efficiently move ions in the direction of negative charge deposition (see, e.g., Fig.~5 in Ref.~\cite{richmondIonosphericElectrodynamicsTutorial2000}).

The above paragraphs describes the energy source for the mechanism suggested by Ref.~\cite{lyatsky_central_1997}, and the resulting fast motions are found in Ref.~\cite{ivarsen_transient_2025}, reporting recent radar measurements of fast-moving electric field structures in the E-region ionosphere. Figure~\ref{fig:cartoon}b) presents a schematic of the process, and thereby summarizes the results of Ref.~\cite{ivarsen_transient_2025} and the present article: equatorward of the cusp and poleward of diffuse aurorae, a series of super-fast turbulent Hall flow channels briefly appeared during two major geomagnetic storms. Given the proximity to the ion precipitation inside the cusp, an experiment that superficially resembles a parallel-plate capacitor emerges. Aided by increased plasma convection caused by dayside reconnection, the action may in the end transport plasma into the polar cap, thereby seeding polar cap patches eastward of the cusp.

Since dynamic and active aurorae dissipate energy, partially through turbulence, the question of what powers them strikes at the heart of the ionosphere-magnetosphere coupling; the greater energy flow from the solar wind down to Earth's dense resistor of an ionosphere. Whereas traditional poleward-moving auroral forms are powered directly by the coupling that takes place in the cusp-region, that of the ionosphere and the solar wind, we here evoke a mechanism involving oscillatory magnetic energy and the particle populations near the magnetospheric equator. In particular, whistler-mode chorus waves are the ultimate source of free energy in the system under study (a bold claim that is substantiated in a companion paper \cite{ivarsen_transient_2025}).

The core dataset used in the present article consists of coherent radar echoes from the unstable E-region, also called the radar aurora \cite{hysellRadarAurora2015}. In the companion paper, Ref.~\cite{ivarsen_transient_2025}, we present in detail observations of dynamic radar aurorae that would seem to drift poleward but mostly eastward, at a location equatorward of the cusp and on \textit{closed} magnetic field-lines, which we measured using the \textsc{icebear} radar. \textsc{icebear} is a coherent-scatter radar \cite{huyghebaertICEBEARAlldigitalBistatic2019}  whose signal reflects off small (3~meters) Farley-Buneman Waves, unstable structures excited by the relative motion between electrons and ions \cite{farleyPlasmaInstabilityResulting1963,bunemanExcitationFieldAligned1963}. The small-scale turbulence dissipates fast, giving the radar aurora an ephemeral quality \cite{prikryl_doppler_1988,prikryl_evidence_1990}. Ref.~\cite{ivarsen_point-cloud_2024} exploits this quality to identify and track moving clusters of echoes, the apparent motion of which act as a measure of the ionospheric electric field \cite{ivarsen_deriving_2024}.

Based on such electric field observations through two two-hour intervals, occuring during the 23 April 2023 and 10 May 2024 geomagnetic storms, the present article substantiates a series of transient, turbulent Hall currents that formed equatorward of the cusp, and for the 23 April 2023 storm, we demonstrated in a companion paper that those motions were observed on the poleward side of dayside diffuse aurorae, and that the radar observations were strongly correlated with chorus-wave-driven particle precipitation, observed near the equatorial magnetopause \cite{ivarsen_transient_2025}.

\begin{figure}
    \centering
    \includegraphics[width=0.5\textwidth]{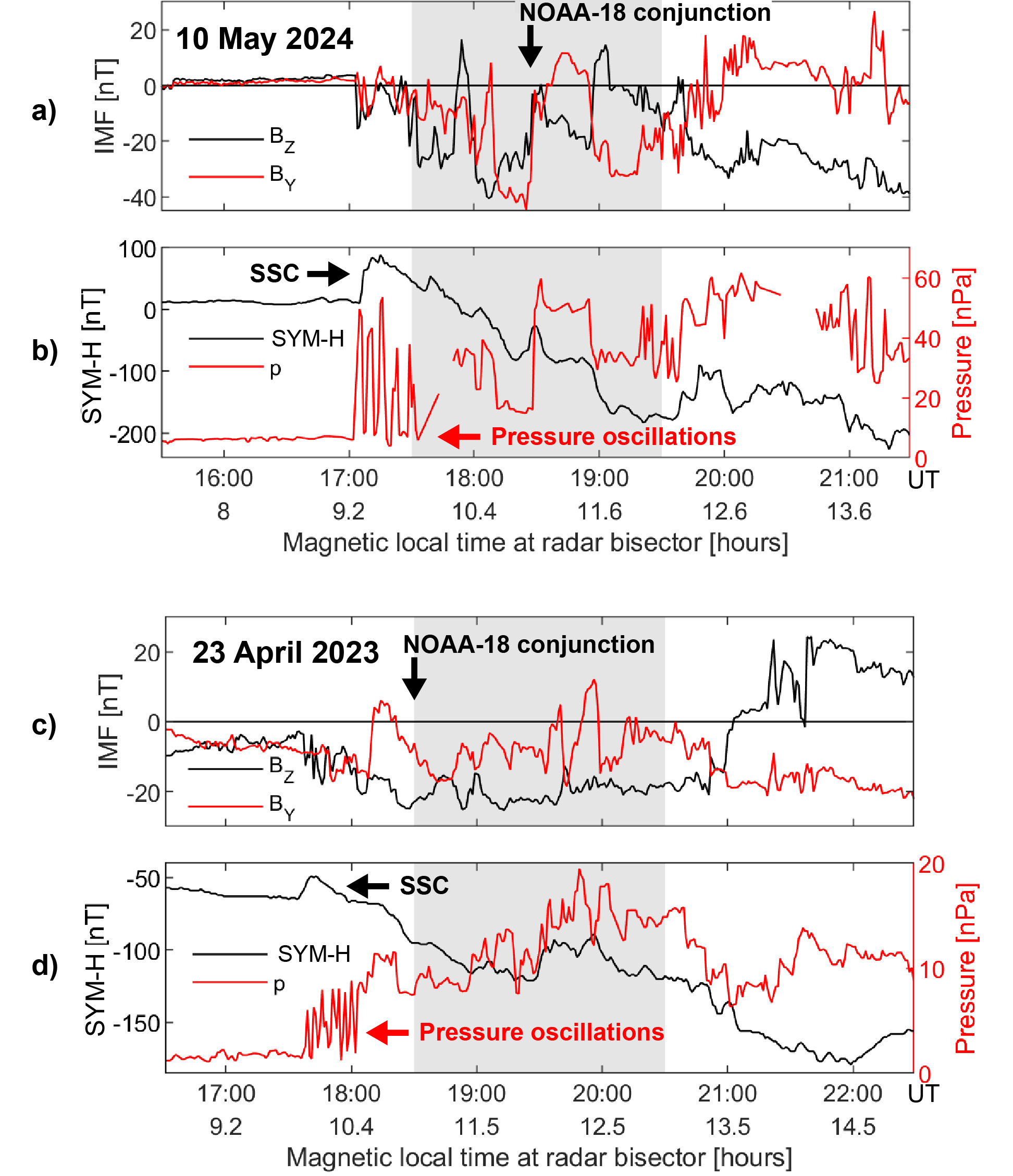}
    \caption{Solar wind parameters and geomagnetic activity index values for the period leading up to two major geomagnetic storms, occurring on 10 May 2024 (panels a, b) and 23 April 2023 (c, d). A shaded gray area denotes the duration of the two events under study. Panels a) and c) show the interplanetary magnetic field $B_Z$ (black) and $B_Y$ (red) components timeshifted to the bowshock \cite{papitashviliOMNIHourlyData2020}. Panels b) and d) show the Sym-H geomagnetic storm index (black, left axis) and the solar wind dynamic pressure (red, right axis). Various features are annotated with arrows (SSC stands for sudden storm commencement).}
    \label{fig:indices}
\end{figure}

\begin{figure}
    \centering
    \includegraphics[width=0.5\textwidth]{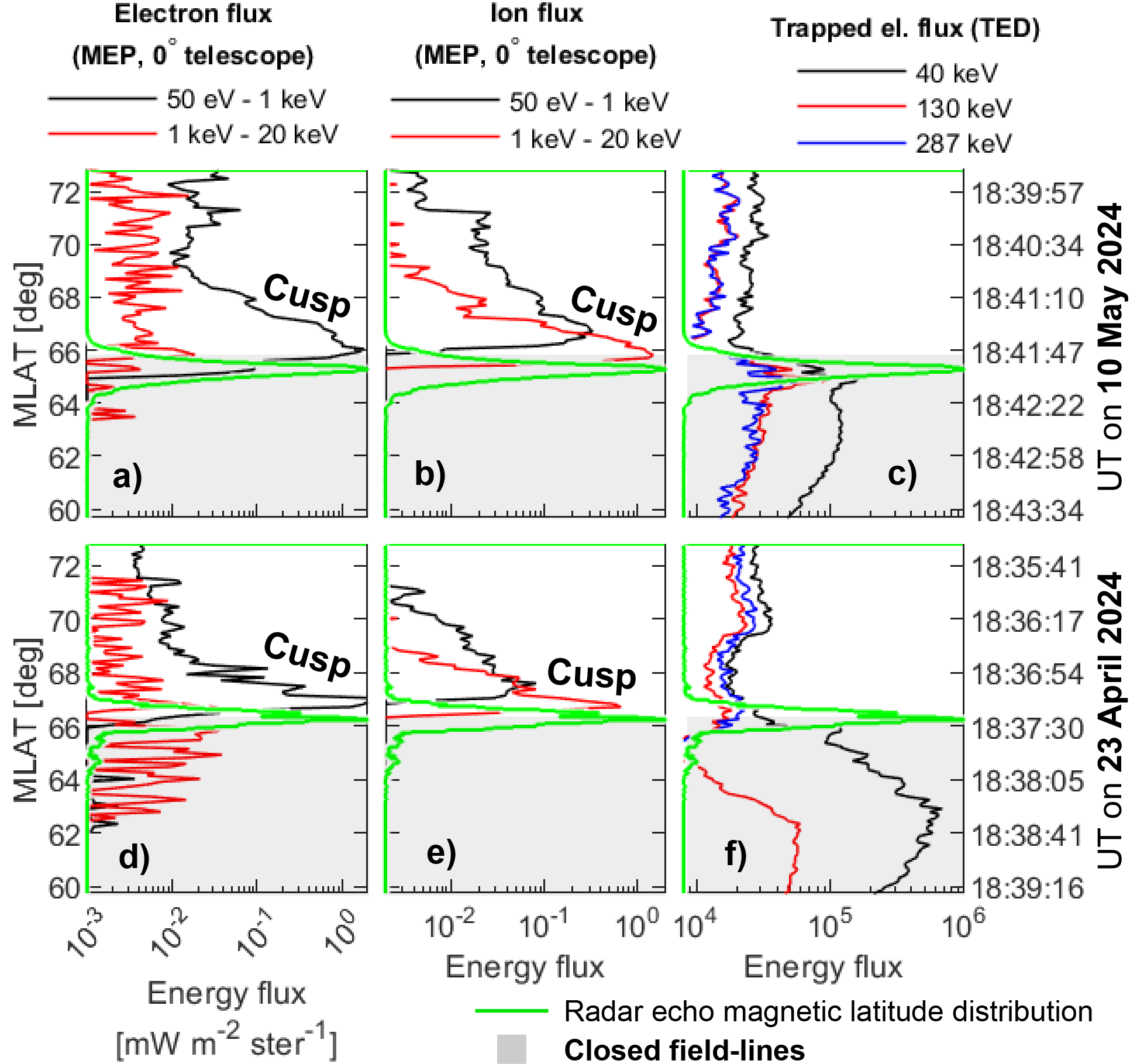}
    \caption{Observations made by the NOAA-18 spacecraft \cite{davis_history_2007} of precipitating electrons (a, d), ions (b, e),  using the TED 0$^\circ$ telescope, as well as trapped electrons (c, f), using the MEPED 30$^\circ$ telescope \cite{evans_polar_2000}, during two space-ground conjunctions; one occurring on 10 May 2024 (a--c) and one occurring on 23 April 2023 (d--f). Magnetic latitude (calculating using AACGM \cite{bakerNewMagneticCoordinate1989}) is shown on the left $y$-axis, while time in UT is shown on the right $y$-axis. In panels a), b), d), and e), black and red lines indicate low- and high-energy particle fluxes respectively, while panels c) and f) show three distinct high-energy fluxes. In all panels, the latitudinal distribution of \textsc{icebear} echoes (with locations traced along Earth's magnetic field-lines) are shown with a green line. A grey shaded region indicates where Earth's magnetic field-lines are inferred to be \textit{closed}.}
    \label{fig:noaa}
\end{figure}

\begin{figure*}
    \centering
    \includegraphics[width=\textwidth]{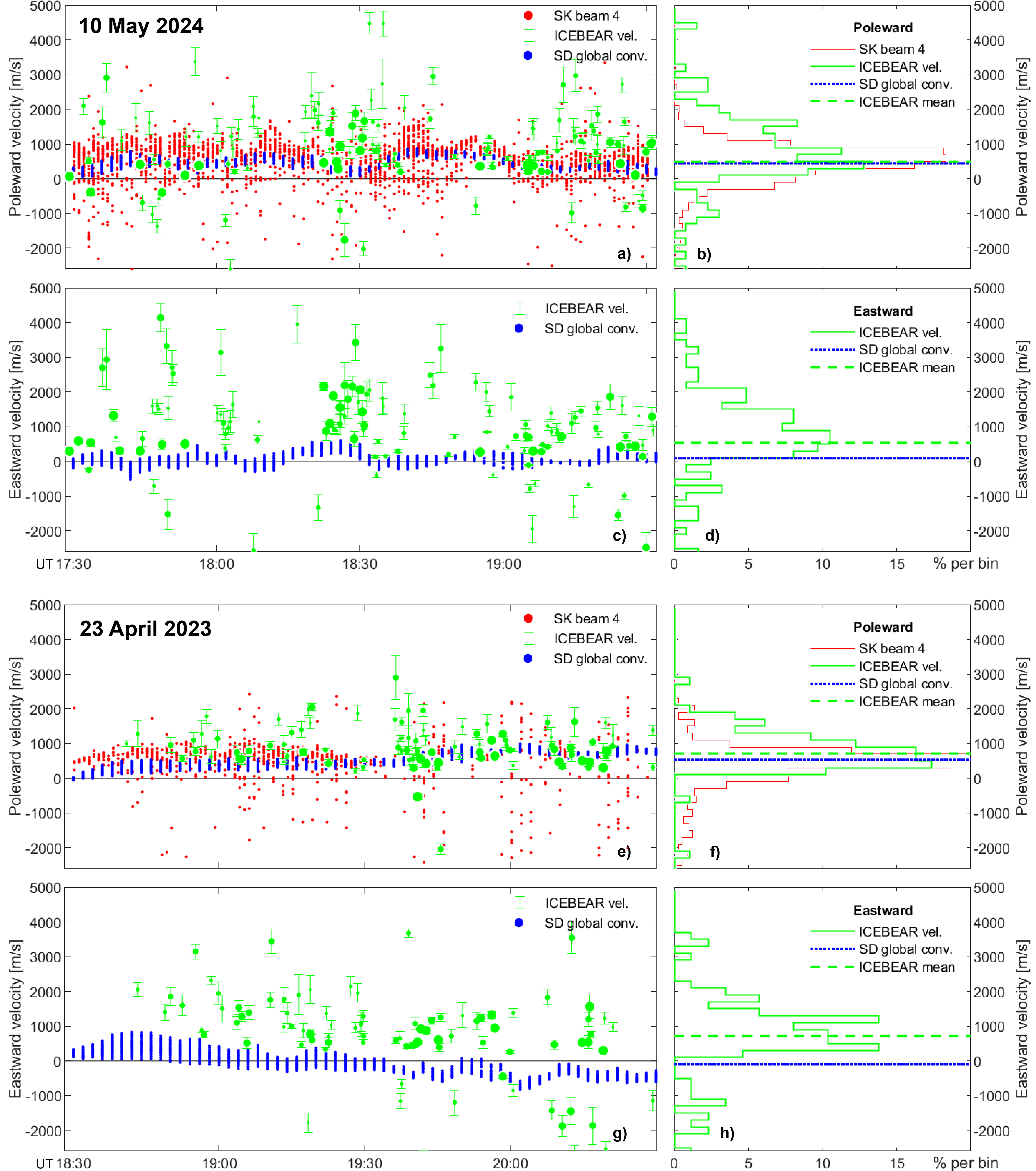}
    \caption{\textbf{\textsc{icebear} cluster motions (green) compared to F-region velocities from Super\textsc{darn}, both observations (red) and model-derived (blue).} The leftmost column shows the temporal development in velocities while the rightmost column shows histograms; the four top panels show the 10 May 2024 event, while the bottom four panels show the 23 April 2023 event. Tracked \textsc{icebear} echo clusters were selected for containing a minimum of 300 echoes, a minimum duration of 6~seconds, as well as variability (68-percent confidence intervals of the linear fits of the echo cluster motion measurements) that did not exceed 2/3 of the cluster speed itself (variability is shown by green errorbars). For poleward motions (panels a, b, e, and f), Doppler shifts measured by the Saskatoon Super\textsc{darn} radar are shown in red (using beam 4, which is pointing in the direction of the geomagnetic north pole).}
    \label{fig:velocities0}
\end{figure*}

\section{Results}

Figure~\ref{fig:indices} elucidates the state of the solar wind prior to the 23 April 2023 (panels a, b) and 10 May 2024 (c, d) storms. In both cases, our observations occurred during geomagnetic noon in Western Canada, in full daylight, coinciding with the two storms' main phases. Furthermore, the two events share several key characteristics; a clear sudden storm commencement coincided with intense pressure oscillations, indicative of non-linear processes near the magnetosheath, possibly triggering bow shock ripples and high-speed jets \cite{li_bow_2024}, sources of free energy for the dayside ionosphere on closed field-lines \cite{kramer_jets_2024}. In both events, the pressure oscillations are followed by a severe southward turn of the interplanetary magnetic field, down to $B_Z<-25$~nT and $B_Z<-40$~nT respectively. In both cases, the southward shift in the $B_Z$ components was severe enough that the open-closed field-line boundary and the dayside auroral region migrated equatorward into the field-of-view of \textsc{icebear} and the Saskatoon Super\textsc{darn} radar.

The similarities between the two events under study do not end there. Figure~\ref{fig:noaa} presents two fortuitous space-ground conjunctions between the United States' National Oceanic and Atmospheric Administration
(NOAA) polar orbiting climate satellite,  reveal the exact location of the ionospheric cusp and the open-closed field-line boundary, which is compared to the distribution of observed radar echoes -- for both the events under study. The satellite is in a polar orbit (98.74$^\circ$ inclination), and is equipped with rudimentary particle detectors that allow for an estimate of the precipitating flux of low- and high-energy electrons and ions, as well as the trapped particle population in the topside F-region. In both the events under analysis, we observe a sharp increase in the low-energy electron energy flux (black line in panels a and d), coincident with a localized, strong ion energy flux (panels b and e), a characteristic signature of the ionospheric cusp \cite{newellCuspCleftBoundary1988,ivarsenGNSSScintillationsCusp2023}. And we here note a tendency for the most energetic and numerous protons usually being located on the equatorward edge of the cusp during southward interplanetary magnetic field configuration, conforming to expectations for a southward interplanetary magnetic field configuration (compare, say, Figure~\ref{fig:noaa}b,~e with Figure~2 in Ref.~\cite{asai_latitudinal_2005}).

Equatorward of this position, the abrupt change in the trapped electron flux (panels c and f) is a signature of the open-closed field-line boundary (OCB) \cite{jinGPSScintillationsAssociated2017}. In all panels of Figure~\ref{fig:noaa}, the latitudinal distribution of \textsc{icebear} echoes (green line) confirm that the radar echoes originate just equatorward of the cusp, on or near to the open-closed field-line boundary.

Figure~\ref{fig:velocities0} summarizes the various velocities measured during the two events, with poleward velocities shown in panels a), b), e), and f), while eastward velocities are shown in panels c), d), g), and h). Measurements from Super\textsc{darn} are plotted in red and blue, utilizing (F-region) data both from the local radar and inferred from the global convection model respectively. In green circle datapoints we show the \textsc{icebear} velocities, which are the  motions of E-region radar aurorae, as they appear and disappear inside of the radar's field-of-view.  The rightmost panels show histograms of the various velocities posted in the leftmost panels. 

It is important to note that the bulk motion of irregular structures in the E-region has long been poorly understood. Individual turbulent structures generally do not move faster than the local ion sound speed, that is, the zero-growth condition for the Farley-Buneman instability \cite{nielsen_first_1983,fosterSimultaneousObservationsEregion2000}. A value of around 300~m/s -- 600~m/s is favoured. A recent paper, Ref.~\cite{ivarsen_deriving_2024}, interprets the much faster \textit{apparent} motion of radar aurorae as being caused by the ephemeral nature of small-scale plasma turbulence: when the instability drivers  (electric field enhancements) move, new turbulent waves are continuously excited along their paths, with each turbulent wave quickly saturating and dissipating. We reiterate, the apparent radar motions, or `echo bulk velocities', act as proxy measurements of the ionospheric electric field \cite{ivarsen_point-cloud_2024,ivarsen_deriving_2024}.

In panels a--b) and e--f) of Figure~\ref{fig:velocities0}, the poleward component of the various measured and estimated velocities, the distribution of \textsc{icebear} velocities matches that of the local Super\textsc{darn} radar, though there is a much greater spread in the former, and the weighted mean echo velocity (weighted by the number of echoes per tracked cluster) equals the average convection speed inside \textsc{icebear}'s field-of-view, the blue line in panels a) and b) of Figure~\ref{fig:velocities0}. However, this agreement is not the case for the eastward motions. Here, \textsc{icebear} consistently see much faster velocities, with individually tracked clusters measuring almost 5000~m/s, in a region where the Super\textsc{darn} global data assimilation model predicts a low or even negative (westward) convection velocity.

The direct implication of Refs.~\cite{ivarsen_point-cloud_2024,ivarsen_deriving_2024} are that the disproportionately fast \textsc{icebear} velocities that we present in Figure~\ref{fig:velocities0} are caused by highly localized electric field structures (see, in particular, Fig.~5 in Ref.~\cite{ivarsen_point-cloud_2024} and Figs.~3 and 4 in Ref.~\cite{ivarsen_deriving_2024}), capable of saturating the Farley-Buneman instability, that are ostensibly missed by the model-based estimation by Super\textsc{darn}.

\begin{figure}
    \centering
    \includegraphics[width=.5\textwidth]{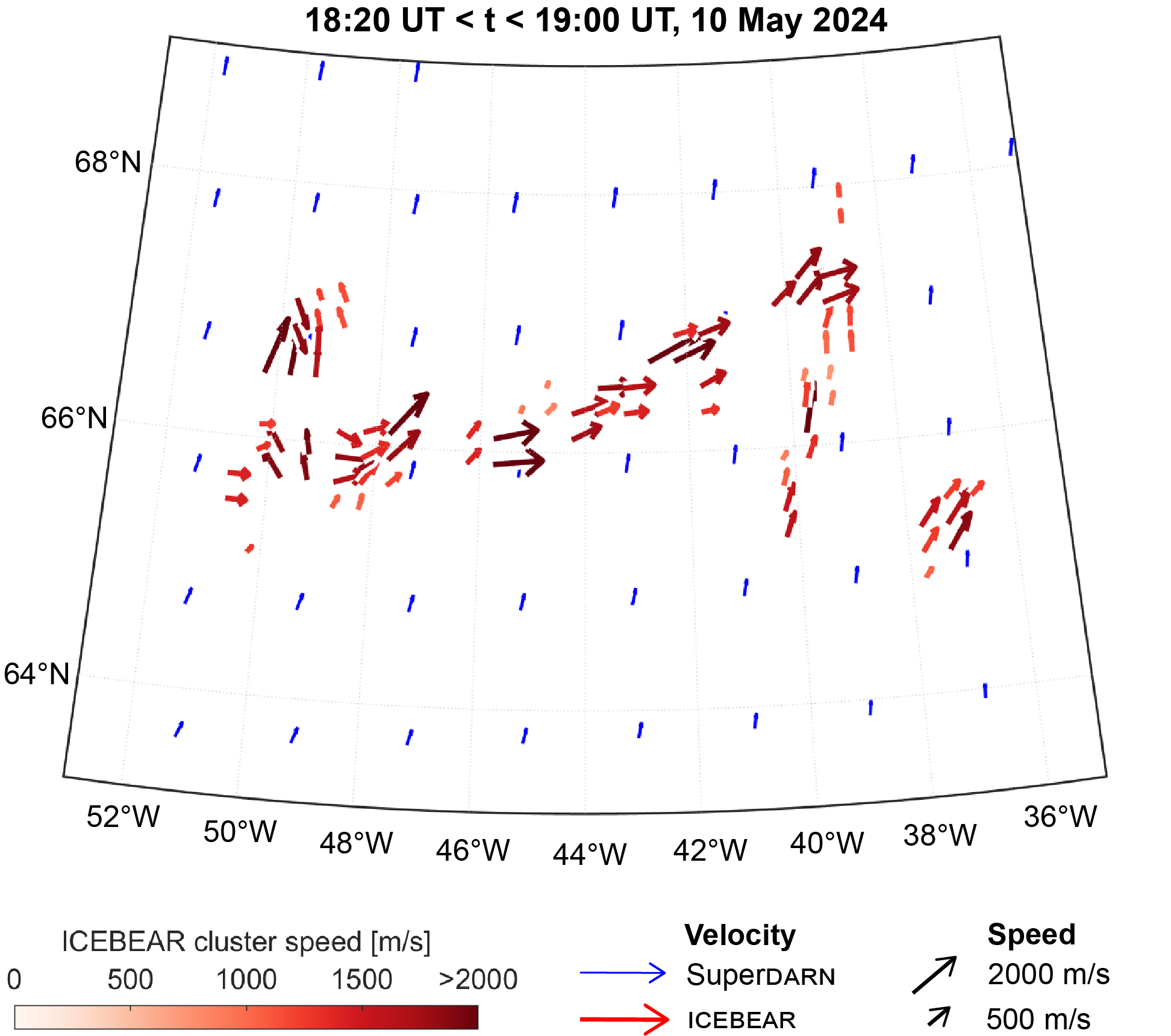}
    \caption{\textsc{icebear} echo velocities in geospace are shown as red arrows (speed indicated with a colorscale), and Super\textsc{darn} global convection velocity are shown as blue arrows, for a 40~minute interval on 10 May 2024. See Figure~\ref{fig:velocities0}a--d) for a detailed description of the directions and magnitudes of the measured and estimated velocities.}
    \label{fig:main}
\end{figure}

Figure~\ref{fig:main} shows the situation in spatial terms, with arrows representing velocities in geomagnetic coordinates. The Super\textsc{darn}-estimated convection pattern is represented by thin blue arrows, while the \textsc{icebear} cluster velocities are represented by red arrows (magnitude shown with a colorscale), for the interval between 18:20~UT -- 19:00~UT on 10 May 2024 (for a similar plot pertaining to 23 April 2023, see Fig.~5 in the companion paper, Ref.~\cite{ivarsen_transient_2025}). Figure~\ref{fig:main} shows very clearly that the \textsc{icebear}-derived velocities in the E-region are heavily skewed eastward, considerably faster than the convection would imply.

\subsubsection*{Description of geospace}

We are in a position to present a sufficiently lucid description of two very similar, but also very surprising datasets. As is shown in Figure~\ref{fig:indices}, storm sudden commencement-events coincided with the onset of rapid, high-amplitude fluctuations in the solar wind dynamic pressure; consistent with the trigger of high-speed jets in the magnetosheath \cite{archer_direct_2019,kramer_jets_2024}, which are known to cause aurorae in the noon-sector \cite{qiu_magnetosheath_2024}. In both the events, the solar wind dynamic pressure thereafter went through a manifold-fold increase, causing Earth's dayside magnetosphere to compress, while the interplanetary magnetic field turned severely southward. Dayside compression events modulate wave activity  \cite{li_bow_2024}, leading to the formation of aurorae, \cite{shi_magnetosphere_2020} an outcome also expected from the severely southward interplanetary magnetic field, by which time \textsc{icebear} observed profuse radar aurorae in the E-region on closed magnetic field-lines. The companion paper, Ref.~\cite{ivarsen_transient_2025},  demonstrates that the above expectations were largely all brought to fruition during the 23 April 2023-event, where we establish a clear link between the radar observations and the observed energy flux of precipitating particles as well as wave-particle interactions.

\subsubsection*{Turbulent electrodynamics on closed field-lines}

Figure~\ref{fig:noaa} substantiates the claim that the radar observations, in both the events under study, were made on \textit{closed magnetic field-lines}, on or very close to the open-closed field-line boundary. Furthermore, Figure~\ref{fig:velocities0} makes it very clear that the tracked radar velocities were strongly skewed in the eastward direction and much faster than expected based on the modeled, global F-region convection pattern (on display in Figure~\ref{fig:main}).

These observations do not adhere to expectations for fast motions in and around the cusp (referring to both the proper motion of field-lines and the F-region $\boldsymbol{E}\times\boldsymbol{B}$-drift). Here, traditional poleward-moving auroral forms are excited in the cusp, on open magnetic field-lines, after which they drift poleward with the F-region convection into the polar cap (see, e.g., Fig.~8 in Ref.~\cite{frey_dayside_2019}). Instead, our observations are consistent with the mechanism evoked by Ref.~\cite{lyatsky_central_1997}, where the motion of auroral forms in the noon-sector are powered by the polarization electric field produced by high-energy aurorae on closed magnetic field-lines.

The relative positions of the \textsc{icebear} motions with respect to the cusp, as well as the peculiar cusp precipitation (Figure~\ref{fig:noaa}) is important. As has been shown in the past \cite{asai_latitudinal_2005}, the most numerous and energetic protons here create an equatorward 'edge' of protons. On its immediate equatorward edge we find the \textsc{icebear} echo distributions. Together with regions of dayside diffuse aurorae, which deposits negative charges, a strong electric field forms individually; one pointing away from the ongoing positve charge deposition and one pointing towards the ongoing negative charge deposition. A situation akin to a parallel-plate capacitor momentarily forms, with strong electric fields distributing charges equatorward in the Pedersen direction, the very currents that close the induced current systems. 

It is important here to point out that the electric fields we infer are the very fields that drive the Pedersen currents that are necessitated by magnetohydrodynamic considerations; a perpendicular current of ions moved by strong electric fields.

The same electric fields trigger the Farley-Buneman instability inside the perpendicular current region, in the Hall direction (perpendicular to both the field-lines and the Pedersen direction).  Turbulent electrojets then form between the two regions in geospace, as illustrated in Figure~\ref{fig:cartoon}b). This is the $\boldsymbol{E}\times\boldsymbol{B}$-drift direction, and so the effect would cause field-lines equatorward of the cusp to move eastward, in addition to the poleward component associated with the global convection pattern. 

\section{Discussion}

In the present article, we have showed that fast, eastward motions can be observed equatorward of the cusp, in the E-region and on closed magnetic field-lines. Moreover, these motions are ostensibly much \textit{faster} than the simultaneously observed F-region drifts.Na\"{i}vely, this appears diametrically opposite to the expectations of the E- and F-region interplay. For example, Figure~2 in Ref.~\cite{fosterSimultaneousObservationsEregion2000} shows observations of E-region \emph{Doppler} speeds that were limited to the local ion sound speed (400--600~m/s), while the authors simultaneously observed that the F-region drifts were twice faster. How can we account for this seeming contradiction? The answer is relatively simple. The radar aurora motions seen by \textsc{icebear} track moving electric field source regions, a motion that is naturally unaffected by collisions, and is `frozen into' the geomagnetic field \cite{ivarsen_point-cloud_2024,ivarsen_deriving_2024}

If the moving electric field structures are frozen into the geomagnetic field, Super\textsc{darn} should, during severe storms in geospace, routinely observe these eastward motions.  However, compared to Super\textsc{darn}, \textsc{icebear} operates with 30 and 60 times higher spatial and temporal resolution respectively. This extreme improvement in spatio-temporal resolution is likely to have caused Super\textsc{darn} to have missed the highly transient spikes in the electric field. Such transient, or `spiky' appearances are unsurprising. The perturbed electric fields of the auroral region are, in general, highly localized and dynamic \cite{lanchester_relationship_1996,opgenoorthRegionsStronglyEnhanced1990,tuttle_horizontal_2020,krcelic_variability_2024}, and a considerable Poynting flux lies ``hidden", as it were, in small-scale electric field enhancements \cite{billettHighResolutionPoyntingFlux2022}.   

Increased conductivities in the sunlit ionosphere should amplify (and be amplified by) the number flux of precipitating electrons \cite{liouSeasonalEffectsAuroral2001}, thereby further enhancing the perpendicular electric field at their emission altitude. The elevated conductivities also facilitate the formation of Pedersen currents, which in fact counteract the electric field enhancements, by virtue of neutralizing extant charges.  However, the observed particle precipitation (Fig.~5 in the companion paper, Ref.~\cite{ivarsen_transient_2025}) should drastically increase the ratio of Hall- to Pedersen conductance \cite{seniorRelationshipFieldalignedCurrents1982,hosokawaModulationIonosphericConductance2010}, speaking to the probability of there being observable electrojets. In this telling, the Hall drifts that  we observe in Figure~\ref{fig:velocities0} may have appeared as the poleward portion of the auroral convection vortices, such as in predictions by Hosokawa et al. \cite{hosokawaModulationIonosphericConductance2010,hosokawaLargeFlowShears2013}.

Although a strong, equatorward electric field is expected poleward of diffuse aurorae \cite{ivarsen_characteristic_2025}, this is not among the established expectations for the region equatorward of the cusp. However, inspecting Figure~\ref{fig:noaa}, we do note that peak proton precipitation occurred on the equatorward side during the two events, and such a configuration is indeed established in the literature for southward interplanetary magnetic field configurations \cite{asai_latitudinal_2005}. The results can therefore be taken into account for further proof in support of a distinct equatorward proton 'edge' in the cusp, one that is shifted sufficiently equatorward as to be noticeable by the electric field record in its equatorward vicinity.

The observations in the present paper, illustrated in Figure~\ref{fig:cartoon}b), boil down to what superficially resemble poleward-moving auroral forms, but, instead of moving with the magnetospheric convection on open field-lines from the cusp and into the polar cap, they move mostly eastward, and in one case that eastward motion was driven by the ongoing flux of high-energy, diffuse electrons excited on closed magnetic field-lines \cite{ivarsen_transient_2025}. Transient excursions away from equilibrium, observed equatorward of the cusp, came in the form of fast, eastward motions. In one case, these motions were triggered by bursts of wave-particle activity, though bursts of proton precipitation in the cusp may in principle likewise control the temporal evolution of the signal.

\section{Closing Remarks}

Traditional poleward-moving auroral forms  move on open field-lines  \cite{saundersPolarCuspIonosphere1989,sandholtPolewardMovingAuroral2007} and their effects are felt in the F-region ionosphere \cite{mcwilliams_two-dimensional_2000,oksavikHighresolutionObservationsSmallscale2004,oksavikScintillationLossSignal2015}. In a similar fashion albeit in different ways,  transient, turbulent electrojets can be excited equatorward of the cusp ionosphere during disturbed conditions, powered in part by high-energy diffuse aurorae, a staple phenomenon in the dayside ionosphere during geospace disturbances \cite{nishimura_structures_2013,ni_chorus_2014,feng_lower_2024}. (In an Appendix we elucidate the routine expectations of having dayside, high-energy diffuse particle precipitation equatorward of the cusp.) The observed mechanism (and the theory developed by Ref.~\cite{lyatsky_central_1997}) can potentially be woven into a substantial part of the body of knowledge concerning dynamic dayside aurorae and the magnetosphere-ionosphere coupling. 

Our study is based on an analysis of a two case studies that occurred during two recent, major geospace storms. Similar observations may have eluded previous efforts in part due to the difficulty of observing the E-region ionosphere, which has led to a scarcity in reliable data  \cite{palmrothLowerthermosphereIonosphereLTI2021}. Exacerbating this, \textsc{icebear} is located at auroral latitudes and is not likely to observe the cusp on a regular basis. Further studies of the phenomenon are needed, and in this regard the rich dataset analyzed in the present study may yield additional insights through modeling efforts \cite{wu_penetrating_2022}; especially pertinent will be the changes to Joule heating rates brought on by the introduction of intense Farley-Buneman turbulence  \cite{wiltbergerEffectsElectrojetTurbulence2017,st-mauriceRevisitingBehaviorERegion2021}.

In closing, we have observed a new, dynamic phenomenon near the ionospheric cusp, pertinent to flux of protons through the cusp itself, as well as the transient energy exchanges between the magnetosheath and the dayside ionosphere on closed magnetic field-lines. The fast motions that we have observed share some key characteristics with poleward-moving auroral forms, but the two phenomena have starkly different causes. Naturally, they are not mutually exclusive but can both be taken to account for observations of strong electric field modulations near the ionospheric cusp.\\

\vspace{-12pt}

\section*{ Acknowledgements}
This work is supported in part by the European Space Agency’s Living Planet Grant No. 1000012348 and by the Research Council of Norway (RCN) Grant No. 324859. We acknowledge the support of the Canadian Space Agency (CSA) [20SUGOICEB], the Canada Foundation for Innovation (CFI) John R. Evans Leaders Fund [32117], the Natural Science and Engineering Research Council (NSERC), the Discovery grants program [RGPIN-2019-19135], the Digital Research Alliance of Canada [RRG-4802], and basic research funding from Korea Astronomy and Space Science Institute [KASI2024185002]. \textsc{icebear} 3D echo data for 23 April 2023 is published with DOI \texttt{10.5281/zenodo.14616122}.  Super\textsc{mag} data can be accessed at \texttt{https://supermag.jhuapl.edu/mag/}. Solar wind data from NASA's \textsc{omni} service can be accessed at \url{https://omniweb.gsfc.nasa.gov/}. MFI is thankful to D. Billett \& L. Clausen for stimulating discussions.

We sadly acknowledge the passing of co-author Kathryn McWilliams during the preparation of this manuscript.

\balancecolsandclearpage

\begin{figure}
    \centering
    \includegraphics[width=0.5\textwidth]{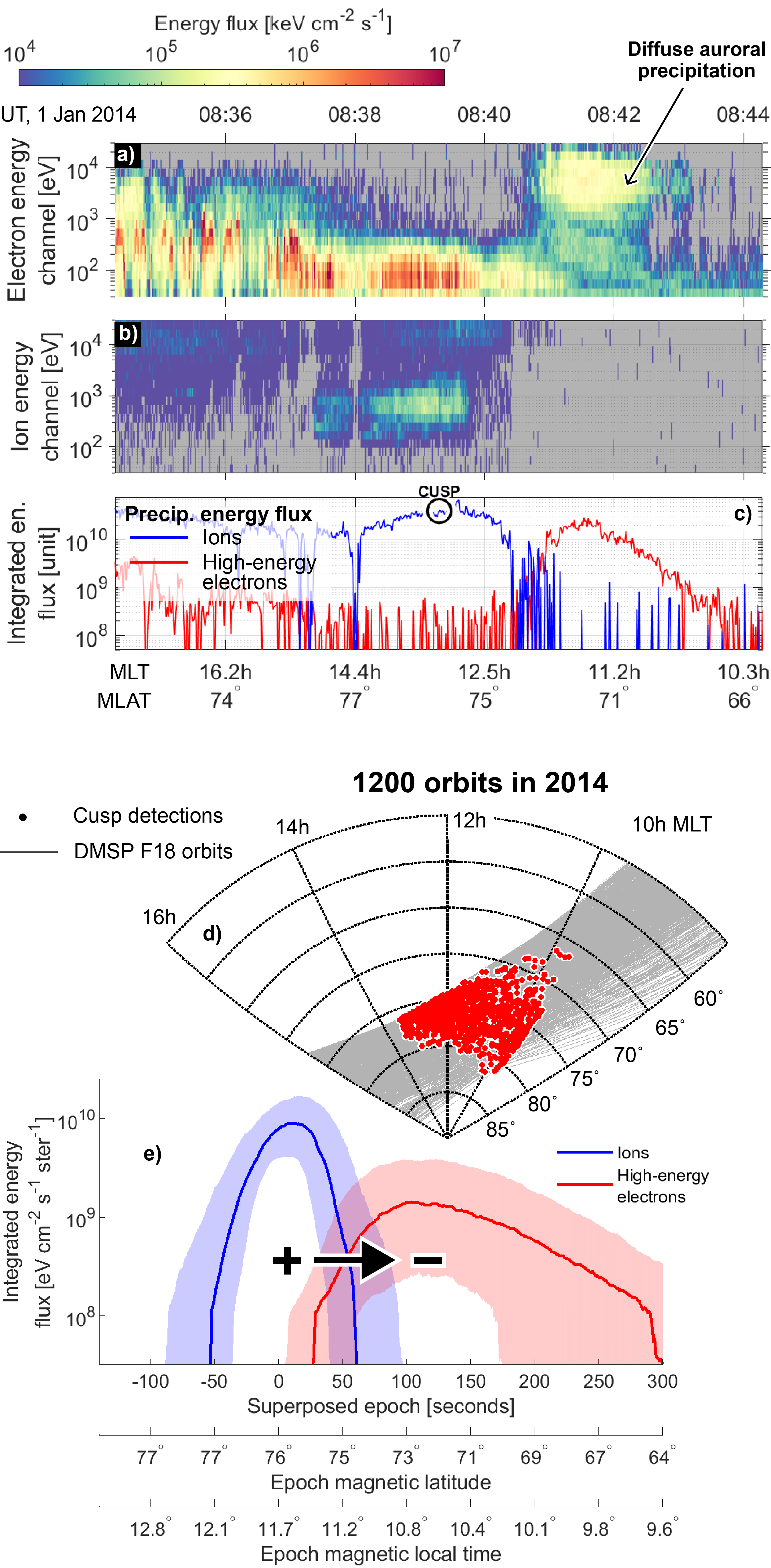}
    \caption{Panels a) and b) show the precipitating electron and ion energy fluxes respectively, during an orbital pass by the DMSP F18 satellite on 1 January 2014, with particle energy along the $y$-axes. Panel c) shows the integrated ion (blue) and the high-energy electron (red) fluxes, with the location of the cusp indicated. Panel d) shows all 1200 orbits made during disturbed conditions (SME-index$~>192~$nT) in 2014, during which the cusp was identified (red circle points, geomagnetic noon is upwards and dawn is to the left). Panel e) shows all 1200 passes through the cusp in a superposed epoch analysis, with the integrated ion and hard electron fluxes as blue and red lines (shaded regions indicate upper and lower quartile distributions).}
    \label{fig:dmsp}
\end{figure}

\section{Appendix}

To provide additional evidence for our explanation beyond the two case studies already covered, we shall introduce a statistical representation of the particle precipitation-landscape near the cusp-region during disturbed conditions. For this purpose, we have turned to the United States' Defense Meteorological Satellite Program (DMSP), a long-lasting mission to study the causes of space weather in the topside ionosphere, whose instrumentation offers precise measurements of the precipitating particle fluxes.

The DMSP satellites operated with helio-synchronous dawn-dusk polar orbits at an altitude of around 840~km, covering most of the northern hemisphere, dayside high-latitude ionosphere. The SSJ instrument consisted of particle detectors that measured the number and energy fluxes of precipitating electrons and ions through 19 energy channels, sensitive to energies between 30 eV and 30 keV, and the instrument operated with a measurement frequency of 1 second \cite{redmonNewDMSPDatabase2017}. We characterize the high-energy electron precipitation by integrating over energy channels between 2~keV and 30~keV, following the method outlined in \cite{redmonNewDMSPDatabase2017}. Furthermore, we classify each precipitating particle as having been directly sampled in the \textit{cusp}, following Ref.~\cite{ivarsenGNSSScintillationsCusp2023} and Ref.~\cite{newellCuspCleftBoundary1988}. Here, a cusp datapoint is defined as having an average precipitating electron energy lower than 220~eV, an average precipitating ion energy higher than 300~eV and lower than 3000~eV, as well as a precipitating electron energy flux through channels 2~keV and 5~keV being lower than 10$^7$~keV~cm$^{-2}$~s$^{-1}$ster$^{-1}$, and the total integrated ion energy flux should exceed  $2\times10^9$~keV~cm$^{-2}$~s$^{-1}$ster$^{-1}$.

After many decades of operation, the DMSP mission is winding down, and there is very limited coverage during the years of our case studies, and so we have analyzed data from 2014, an active year. Figure~\ref{fig:dmsp}a--c) show a pass through the ionospheric cusp by the DMSP F18 satellite on 1 January 2014. The orbit brought the satellite directly through the cusp, after which it traversed a region of characteristic dayside diffuse aurorae (indicated with a black arrow in Figure~\ref{fig:dmsp}a). The two regions are separated in space, with the latter appearing directly equatorward of the former.

Next, we analyze the 1200 orbits that occurred during disturbed conditions, and during which the cusp was successfully identified in the data (Figure~\ref{fig:dmsp}d). Panel e) shows a super-posed epoch analysis of all these orbits, centered on the cusp ($t=0$~seconds), with blue and red lines showing the integrated ion flux and the integrated flux if high-energy ($>2$~keV) electrons respectively. This view of the cusp-region is not new, and is reflected in climatological studies of particle precipitation \cite{newellSeasonalVariationsDiffuse2010,ivarsenGNSSScintillationsCusp2023}.

If we discount the intense, soft electron flux inside the cusp (which will be stopped by Earth's atmosphere far above the E-region \cite{fangParameterizationMonoenergeticElectron2010,ivarsen_what_2024}), the two fluxes in Figure~\ref{fig:dmsp}e) will produce transient space-charge fields that are entirely consistent with the particle precipitation landscape that we are immersed in. While it is highly unlikely that this situation is sufficient to produce measurable electric fields in all the 1200 orbits, the superposed epoch analysis in Figure~\ref{fig:dmsp} presents unequivocal evidence that the conditions necessary for the observations are a routine feature of the noon-sector ionosphere.


\begin{thebibliography}{69}%
\makeatletter
\providecommand \@ifxundefined [1]{%
 \@ifx{#1\undefined}
}%
\providecommand \@ifnum [1]{%
 \ifnum #1\expandafter \@firstoftwo
 \else \expandafter \@secondoftwo
 \fi
}%
\providecommand \@ifx [1]{%
 \ifx #1\expandafter \@firstoftwo
 \else \expandafter \@secondoftwo
 \fi
}%
\providecommand \natexlab [1]{#1}%
\providecommand \enquote  [1]{``#1''}%
\providecommand \bibnamefont  [1]{#1}%
\providecommand \bibfnamefont [1]{#1}%
\providecommand \citenamefont [1]{#1}%
\providecommand \href@noop [0]{\@secondoftwo}%
\providecommand \href [0]{\begingroup \@sanitize@url \@href}%
\providecommand \@href[1]{\@@startlink{#1}\@@href}%
\providecommand \@@href[1]{\endgroup#1\@@endlink}%
\providecommand \@sanitize@url [0]{\catcode `\\12\catcode `\$12\catcode `\&12\catcode `\#12\catcode `\^12\catcode `\_12\catcode `\%12\relax}%
\providecommand \@@startlink[1]{}%
\providecommand \@@endlink[0]{}%
\providecommand \url  [0]{\begingroup\@sanitize@url \@url }%
\providecommand \@url [1]{\endgroup\@href {#1}{\urlprefix }}%
\providecommand \urlprefix  [0]{URL }%
\providecommand \Eprint [0]{\href }%
\providecommand \doibase [0]{https://doi.org/}%
\providecommand \selectlanguage [0]{\@gobble}%
\providecommand \bibinfo  [0]{\@secondoftwo}%
\providecommand \bibfield  [0]{\@secondoftwo}%
\providecommand \translation [1]{[#1]}%
\providecommand \BibitemOpen [0]{}%
\providecommand \bibitemStop [0]{}%
\providecommand \bibitemNoStop [0]{.\EOS\space}%
\providecommand \EOS [0]{\spacefactor3000\relax}%
\providecommand \BibitemShut  [1]{\csname bibitem#1\endcsname}%
\let\auto@bib@innerbib\@empty
\bibitem [{\citenamefont {Dungey}(1961)}]{dungeyInterplanetaryMagneticField1961}%
  \BibitemOpen
  \bibfield  {author} {\bibinfo {author} {\bibfnamefont {J.~W.}\ \bibnamefont {Dungey}},\ }\bibfield  {title} {\bibinfo {title} {Interplanetary {{Magnetic Field}} and the {{Auroral Zones}}},\ }\href {https://doi.org/10.1103/PhysRevLett.6.47} {\bibfield  {journal} {\bibinfo  {journal} {Physical Review Letters}\ }\textbf {\bibinfo {volume} {6}},\ \bibinfo {pages} {47} (\bibinfo {year} {1961})}\BibitemShut {NoStop}%
\bibitem [{\citenamefont {Cowley}(2000)}]{cowleyTUTORIALMagnetosphereIonosphereInteractions2000}%
  \BibitemOpen
  \bibfield  {author} {\bibinfo {author} {\bibfnamefont {S.~W.~H.}\ \bibnamefont {Cowley}},\ }\bibfield  {title} {\bibinfo {title} {{{TUTORIAL}}: {{Magnetosphere-Ionosphere Interactions}}: {{A Tutorial Review}}},\ }\href {https://doi.org/10.1029/GM118p0091} {\bibfield  {journal} {\bibinfo  {journal} {Washington DC American Geophysical Union Geophysical Monograph Series}\ }\textbf {\bibinfo {volume} {118}},\ \bibinfo {pages} {91} (\bibinfo {year} {2000})}\BibitemShut {NoStop}%
\bibitem [{\citenamefont {Saunders}(1989)}]{saundersPolarCuspIonosphere1989}%
  \BibitemOpen
  \bibfield  {author} {\bibinfo {author} {\bibfnamefont {M.}~\bibnamefont {Saunders}},\ }\bibfield  {title} {\bibinfo {title} {The polar cusp ionosphere: A window on solar wind--magnetosphere coupling},\ }\href {https://doi.org/10.1017/S0954102089000313} {\bibfield  {journal} {\bibinfo  {journal} {Antarctic Science}\ }\textbf {\bibinfo {volume} {1}},\ \bibinfo {pages} {193} (\bibinfo {year} {1989})}\BibitemShut {NoStop}%
\bibitem [{\citenamefont {Shepherd}(1979)}]{shepherdDaysideCleftAurora1979}%
  \BibitemOpen
  \bibfield  {author} {\bibinfo {author} {\bibfnamefont {G.~G.}\ \bibnamefont {Shepherd}},\ }\bibfield  {title} {\bibinfo {title} {Dayside cleft aurora and its ionospheric effects},\ }\href {https://doi.org/10.1029/RG017i008p02017} {\bibfield  {journal} {\bibinfo  {journal} {Reviews of Geophysics}\ }\textbf {\bibinfo {volume} {17}},\ \bibinfo {pages} {2017} (\bibinfo {year} {1979})}\BibitemShut {NoStop}%
\bibitem [{\citenamefont {Ogawa}\ \emph {et~al.}(2003)\citenamefont {Ogawa}, \citenamefont {Fujii}, \citenamefont {Buchert}, \citenamefont {Nozawa},\ and\ \citenamefont {Ohtani}}]{ogawaSimultaneousEISCATSvalbard2003}%
  \BibitemOpen
  \bibfield  {author} {\bibinfo {author} {\bibfnamefont {Y.}~\bibnamefont {Ogawa}}, \bibinfo {author} {\bibfnamefont {R.}~\bibnamefont {Fujii}}, \bibinfo {author} {\bibfnamefont {S.~C.}\ \bibnamefont {Buchert}}, \bibinfo {author} {\bibfnamefont {S.}~\bibnamefont {Nozawa}},\ and\ \bibinfo {author} {\bibfnamefont {S.}~\bibnamefont {Ohtani}},\ }\bibfield  {title} {\bibinfo {title} {Simultaneous {{EISCAT Svalbard}} radar and {{DMSP}} observations of ion upflow in the dayside polar ionosphere},\ }\bibfield  {journal} {\bibinfo  {journal} {Journal of Geophysical Research: Space Physics}\ }\textbf {\bibinfo {volume} {108}},\ \href {https://doi.org/10.1029/2002JA009590} {10.1029/2002JA009590} (\bibinfo {year} {2003})\BibitemShut {NoStop}%
\bibitem [{\citenamefont {Frey}\ \emph {et~al.}(2019)\citenamefont {Frey}, \citenamefont {Han}, \citenamefont {Kataoka}, \citenamefont {Lessard}, \citenamefont {Milan}, \citenamefont {Nishimura}, \citenamefont {Strangeway},\ and\ \citenamefont {Zou}}]{frey_dayside_2019}%
  \BibitemOpen
  \bibfield  {author} {\bibinfo {author} {\bibfnamefont {H.~U.}\ \bibnamefont {Frey}}, \bibinfo {author} {\bibfnamefont {D.}~\bibnamefont {Han}}, \bibinfo {author} {\bibfnamefont {R.}~\bibnamefont {Kataoka}}, \bibinfo {author} {\bibfnamefont {M.~R.}\ \bibnamefont {Lessard}}, \bibinfo {author} {\bibfnamefont {S.~E.}\ \bibnamefont {Milan}}, \bibinfo {author} {\bibfnamefont {Y.}~\bibnamefont {Nishimura}}, \bibinfo {author} {\bibfnamefont {R.~J.}\ \bibnamefont {Strangeway}},\ and\ \bibinfo {author} {\bibfnamefont {Y.}~\bibnamefont {Zou}},\ }\bibfield  {title} {\bibinfo {title} {Dayside {{Aurora}}},\ }\href {https://doi.org/10.1007/s11214-019-0617-7} {\bibfield  {journal} {\bibinfo  {journal} {Space Science Reviews}\ }\textbf {\bibinfo {volume} {215}},\ \bibinfo {pages} {51} (\bibinfo {year} {2019})}\BibitemShut {NoStop}%
\bibitem [{\citenamefont {Southwood}\ \emph {et~al.}(1988)\citenamefont {Southwood}, \citenamefont {Farrugia},\ and\ \citenamefont {Saunders}}]{southwoodWhatAreFlux1988}%
  \BibitemOpen
  \bibfield  {author} {\bibinfo {author} {\bibfnamefont {D.~J.}\ \bibnamefont {Southwood}}, \bibinfo {author} {\bibfnamefont {C.~J.}\ \bibnamefont {Farrugia}},\ and\ \bibinfo {author} {\bibfnamefont {M.~A.}\ \bibnamefont {Saunders}},\ }\bibfield  {title} {\bibinfo {title} {What are flux transfer events?},\ }\href {https://doi.org/10.1016/0032-0633(88)90109-2} {\bibfield  {journal} {\bibinfo  {journal} {Planetary and Space Science}\ }\textbf {\bibinfo {volume} {36}},\ \bibinfo {pages} {503} (\bibinfo {year} {1988})}\BibitemShut {NoStop}%
\bibitem [{\citenamefont {Oksavik}\ \emph {et~al.}(2004)\citenamefont {Oksavik}, \citenamefont {Moen},\ and\ \citenamefont {Carlson}}]{oksavikHighresolutionObservationsSmallscale2004}%
  \BibitemOpen
  \bibfield  {author} {\bibinfo {author} {\bibfnamefont {K.}~\bibnamefont {Oksavik}}, \bibinfo {author} {\bibfnamefont {J.}~\bibnamefont {Moen}},\ and\ \bibinfo {author} {\bibfnamefont {H.~C.}\ \bibnamefont {Carlson}},\ }\bibfield  {title} {\bibinfo {title} {High-resolution observations of the small-scale flow pattern associated with a poleward moving auroral form in the cusp},\ }\bibfield  {journal} {\bibinfo  {journal} {Geophysical Research Letters}\ }\textbf {\bibinfo {volume} {31}},\ \href {https://doi.org/10.1029/2004GL019838} {10.1029/2004GL019838} (\bibinfo {year} {2004})\BibitemShut {NoStop}%
\bibitem [{\citenamefont {Fasel}(1995)}]{faselDaysidePolewardMoving1995}%
  \BibitemOpen
  \bibfield  {author} {\bibinfo {author} {\bibfnamefont {G.~J.}\ \bibnamefont {Fasel}},\ }\bibfield  {title} {\bibinfo {title} {Dayside poleward moving auroral forms: {{A}} statistical study},\ }\href {https://doi.org/10.1029/95JA00854} {\bibfield  {journal} {\bibinfo  {journal} {Journal of Geophysical Research: Space Physics}\ }\textbf {\bibinfo {volume} {100}},\ \bibinfo {pages} {11891} (\bibinfo {year} {1995})}\BibitemShut {NoStop}%
\bibitem [{\citenamefont {Lester}(1998)}]{lester_coherent-scatter_1998}%
  \BibitemOpen
  \bibfield  {author} {\bibinfo {author} {\bibfnamefont {M.}~\bibnamefont {Lester}},\ }\bibfield  {title} {\bibinfo {title} {Coherent-{{Scatter Radar Studies}} of the {{Dayside Cusp}}},\ }in\ \href {https://doi.org/10.1007/978-94-011-5214-3_17} {\emph {\bibinfo {booktitle} {Polar {{Cap Boundary Phenomena}}}}},\ \bibinfo {editor} {edited by\ \bibinfo {editor} {\bibfnamefont {J.}~\bibnamefont {Moen}}, \bibinfo {editor} {\bibfnamefont {A.}~\bibnamefont {Egeland}},\ and\ \bibinfo {editor} {\bibfnamefont {M.}~\bibnamefont {Lockwood}}}\ (\bibinfo  {publisher} {Springer Netherlands},\ \bibinfo {address} {Dordrecht},\ \bibinfo {year} {1998})\ pp.\ \bibinfo {pages} {219--232}\BibitemShut {NoStop}%
\bibitem [{\citenamefont {Sandholt}\ and\ \citenamefont {Farrugia}(2007)}]{sandholtPolewardMovingAuroral2007}%
  \BibitemOpen
  \bibfield  {author} {\bibinfo {author} {\bibfnamefont {P.~E.}\ \bibnamefont {Sandholt}}\ and\ \bibinfo {author} {\bibfnamefont {C.~J.}\ \bibnamefont {Farrugia}},\ }\bibfield  {title} {\bibinfo {title} {Poleward moving auroral forms ({{PMAFs}}) revisited: Responses of aurorae, plasma convection and {{Birkeland}} currents in the pre- and postnoon sectors under positive and negative {{IMF}} {{{\emph{B}}}}{\emph{{\textsubscript{y}}}} conditions},\ }\href {https://doi.org/10.5194/angeo-25-1629-2007} {\bibfield  {journal} {\bibinfo  {journal} {Annales Geophysicae}\ }\textbf {\bibinfo {volume} {25}},\ \bibinfo {pages} {1629} (\bibinfo {year} {2007})}\BibitemShut {NoStop}%
\bibitem [{\citenamefont {Oksavik}\ \emph {et~al.}(2015)\citenamefont {Oksavik}, \citenamefont {{van der Meeren}}, \citenamefont {Lorentzen}, \citenamefont {Baddeley},\ and\ \citenamefont {Moen}}]{oksavikScintillationLossSignal2015}%
  \BibitemOpen
  \bibfield  {author} {\bibinfo {author} {\bibfnamefont {K.}~\bibnamefont {Oksavik}}, \bibinfo {author} {\bibfnamefont {C.}~\bibnamefont {{van der Meeren}}}, \bibinfo {author} {\bibfnamefont {D.~A.}\ \bibnamefont {Lorentzen}}, \bibinfo {author} {\bibfnamefont {L.~J.}\ \bibnamefont {Baddeley}},\ and\ \bibinfo {author} {\bibfnamefont {J.}~\bibnamefont {Moen}},\ }\bibfield  {title} {\bibinfo {title} {Scintillation and loss of signal lock from poleward moving auroral forms in the cusp ionosphere},\ }\href {https://doi.org/10.1002/2015JA021528} {\bibfield  {journal} {\bibinfo  {journal} {Journal of Geophysical Research: Space Physics}\ }\textbf {\bibinfo {volume} {120}},\ \bibinfo {pages} {9161} (\bibinfo {year} {2015})}\BibitemShut {NoStop}%
\bibitem [{\citenamefont {Milan}\ \emph {et~al.}(2000)\citenamefont {Milan}, \citenamefont {Lester}, \citenamefont {Cowley},\ and\ \citenamefont {Brittnacher}}]{milanConvectionAuroralResponse2000}%
  \BibitemOpen
  \bibfield  {author} {\bibinfo {author} {\bibfnamefont {S.~E.}\ \bibnamefont {Milan}}, \bibinfo {author} {\bibfnamefont {M.}~\bibnamefont {Lester}}, \bibinfo {author} {\bibfnamefont {S.~W.~H.}\ \bibnamefont {Cowley}},\ and\ \bibinfo {author} {\bibfnamefont {M.}~\bibnamefont {Brittnacher}},\ }\bibfield  {title} {\bibinfo {title} {Convection and auroral response to a southward turning of the {{IMF}}: {{Polar UVI}}, {{CUTLASS}}, and {{IMAGE}} signatures of transient magnetic flux transfer at the magnetopause},\ }\href {https://doi.org/10.1029/2000JA900022} {\bibfield  {journal} {\bibinfo  {journal} {Journal of Geophysical Research: Space Physics}\ }\textbf {\bibinfo {volume} {105}},\ \bibinfo {pages} {15741} (\bibinfo {year} {2000})}\BibitemShut {NoStop}%
\bibitem [{\citenamefont {McWilliams}\ \emph {et~al.}(2000)\citenamefont {McWilliams}, \citenamefont {Yeoman},\ and\ \citenamefont {Cowley}}]{mcwilliams_two-dimensional_2000}%
  \BibitemOpen
  \bibfield  {author} {\bibinfo {author} {\bibfnamefont {K.~A.}\ \bibnamefont {McWilliams}}, \bibinfo {author} {\bibfnamefont {T.~K.}\ \bibnamefont {Yeoman}},\ and\ \bibinfo {author} {\bibfnamefont {S.~W.~H.}\ \bibnamefont {Cowley}},\ }\bibfield  {title} {\bibinfo {title} {Two-dimensional electric field measurements in the ionospheric footprint of a flux transfer event},\ }\href {https://doi.org/10.1007/s00585-001-1584-2} {\bibfield  {journal} {\bibinfo  {journal} {Annales Geophysicae}\ }\textbf {\bibinfo {volume} {18}},\ \bibinfo {pages} {1584} (\bibinfo {year} {2000})}\BibitemShut {NoStop}%
\bibitem [{\citenamefont {Sandholt}(1991)}]{sandholtAuroralElectrodynamicsCusp1991}%
  \BibitemOpen
  \bibfield  {author} {\bibinfo {author} {\bibfnamefont {P.~E.}\ \bibnamefont {Sandholt}},\ }\bibfield  {title} {\bibinfo {title} {Auroral electrodynamics at the cusp/cleft poleward boundary during northward interplanetary magnetic field},\ }\href {https://doi.org/10.1029/91GL00849} {\bibfield  {journal} {\bibinfo  {journal} {Geophysical Research Letters}\ }\textbf {\bibinfo {volume} {18}},\ \bibinfo {pages} {805} (\bibinfo {year} {1991})}\BibitemShut {NoStop}%
\bibitem [{\citenamefont {Kozlovsky}\ and\ \citenamefont {Kangas}(2002)}]{kozlovsky_motion_2002}%
  \BibitemOpen
  \bibfield  {author} {\bibinfo {author} {\bibfnamefont {A.}~\bibnamefont {Kozlovsky}}\ and\ \bibinfo {author} {\bibfnamefont {J.}~\bibnamefont {Kangas}},\ }\bibfield  {title} {\bibinfo {title} {Motion and origin of noon high-latitude poleward moving auroral arcs on closed magnetic field lines},\ }\href {https://doi.org/10.1029/2001JA900145} {\bibfield  {journal} {\bibinfo  {journal} {Journal of Geophysical Research: Space Physics}\ }\textbf {\bibinfo {volume} {107}},\ \bibinfo {pages} {SMP 1} (\bibinfo {year} {2002})}\BibitemShut {NoStop}%
\bibitem [{\citenamefont {Wu}\ \emph {et~al.}(2020)\citenamefont {Wu}, \citenamefont {Mende},\ and\ \citenamefont {Frey}}]{wu_simultaneous_2020}%
  \BibitemOpen
  \bibfield  {author} {\bibinfo {author} {\bibfnamefont {Y.-J.~J.}\ \bibnamefont {Wu}}, \bibinfo {author} {\bibfnamefont {S.~B.}\ \bibnamefont {Mende}},\ and\ \bibinfo {author} {\bibfnamefont {H.~U.}\ \bibnamefont {Frey}},\ }\bibfield  {title} {\bibinfo {title} {Simultaneous {{Observations}} of {{Poleward-Moving Auroral Forms}} at the {{Equatorward}} and {{Poleward Boundaries}} of the {{Auroral Oval}} in {{Antarctica}}},\ }\href {https://doi.org/10.1029/2019JA027646} {\bibfield  {journal} {\bibinfo  {journal} {Journal of Geophysical Research: Space Physics}\ }\textbf {\bibinfo {volume} {125}},\ \bibinfo {pages} {e2019JA027646} (\bibinfo {year} {2020})}\BibitemShut {NoStop}%
\bibitem [{\citenamefont {Lyatsky}\ and\ \citenamefont {Sibeck}(1997)}]{lyatsky_central_1997}%
  \BibitemOpen
  \bibfield  {author} {\bibinfo {author} {\bibfnamefont {W.}~\bibnamefont {Lyatsky}}\ and\ \bibinfo {author} {\bibfnamefont {D.}~\bibnamefont {Sibeck}},\ }\bibfield  {title} {\bibinfo {title} {Central plasma sheet disruption and the formation of dayside poleward moving auroral events},\ }\href {https://doi.org/10.1029/97JA01402} {\bibfield  {journal} {\bibinfo  {journal} {Journal of Geophysical Research: Space Physics}\ }\textbf {\bibinfo {volume} {102}},\ \bibinfo {pages} {17625} (\bibinfo {year} {1997})}\BibitemShut {NoStop}%
\bibitem [{\citenamefont {Lyatsky}\ \emph {et~al.}(2016)\citenamefont {Lyatsky}, \citenamefont {Pollock}, \citenamefont {Goldstein}, \citenamefont {Lyatskaya},\ and\ \citenamefont {Avanov}}]{lyatsky_penetration_2016}%
  \BibitemOpen
  \bibfield  {author} {\bibinfo {author} {\bibfnamefont {W.}~\bibnamefont {Lyatsky}}, \bibinfo {author} {\bibfnamefont {C.}~\bibnamefont {Pollock}}, \bibinfo {author} {\bibfnamefont {M.~L.}\ \bibnamefont {Goldstein}}, \bibinfo {author} {\bibfnamefont {S.}~\bibnamefont {Lyatskaya}},\ and\ \bibinfo {author} {\bibfnamefont {L.}~\bibnamefont {Avanov}},\ }\bibfield  {title} {\bibinfo {title} {Penetration of magnetosheath plasma into dayside magnetosphere: 1. {{Density}}, velocity, and rotation},\ }\href {https://doi.org/10.1002/2015JA022119} {\bibfield  {journal} {\bibinfo  {journal} {Journal of Geophysical Research: Space Physics}\ }\textbf {\bibinfo {volume} {121}},\ \bibinfo {pages} {7699} (\bibinfo {year} {2016})}\BibitemShut {NoStop}%
\bibitem [{\citenamefont {Spasojevic}\ and\ \citenamefont {Inan}(2010)}]{spasojevic_drivers_2010}%
  \BibitemOpen
  \bibfield  {author} {\bibinfo {author} {\bibfnamefont {M.}~\bibnamefont {Spasojevic}}\ and\ \bibinfo {author} {\bibfnamefont {U.~S.}\ \bibnamefont {Inan}},\ }\bibfield  {title} {\bibinfo {title} {Drivers of chorus in the outer dayside magnetosphere},\ }\bibfield  {journal} {\bibinfo  {journal} {Journal of Geophysical Research: Space Physics}\ }\textbf {\bibinfo {volume} {115}},\ \href {https://doi.org/10.1029/2009JA014452} {10.1029/2009JA014452} (\bibinfo {year} {2010})\BibitemShut {NoStop}%
\bibitem [{\citenamefont {Nishimura}\ \emph {et~al.}(2013)\citenamefont {Nishimura}, \citenamefont {Bortnik}, \citenamefont {Li}, \citenamefont {Thorne}, \citenamefont {Ni}, \citenamefont {Lyons}, \citenamefont {Angelopoulos}, \citenamefont {Ebihara}, \citenamefont {Bonnell}, \citenamefont {Le~Contel},\ and\ \citenamefont {Auster}}]{nishimura_structures_2013}%
  \BibitemOpen
  \bibfield  {author} {\bibinfo {author} {\bibfnamefont {Y.}~\bibnamefont {Nishimura}}, \bibinfo {author} {\bibfnamefont {J.}~\bibnamefont {Bortnik}}, \bibinfo {author} {\bibfnamefont {W.}~\bibnamefont {Li}}, \bibinfo {author} {\bibfnamefont {R.~M.}\ \bibnamefont {Thorne}}, \bibinfo {author} {\bibfnamefont {B.}~\bibnamefont {Ni}}, \bibinfo {author} {\bibfnamefont {L.~R.}\ \bibnamefont {Lyons}}, \bibinfo {author} {\bibfnamefont {V.}~\bibnamefont {Angelopoulos}}, \bibinfo {author} {\bibfnamefont {Y.}~\bibnamefont {Ebihara}}, \bibinfo {author} {\bibfnamefont {J.~W.}\ \bibnamefont {Bonnell}}, \bibinfo {author} {\bibfnamefont {O.}~\bibnamefont {Le~Contel}},\ and\ \bibinfo {author} {\bibfnamefont {U.}~\bibnamefont {Auster}},\ }\bibfield  {title} {\bibinfo {title} {Structures of dayside whistler-mode waves deduced from conjugate diffuse aurora},\ }\href {https://doi.org/10.1029/2012JA018242} {\bibfield  {journal} {\bibinfo  {journal} {Journal of Geophysical Research: Space Physics}\ }\textbf {\bibinfo {volume}
  {118}},\ \bibinfo {pages} {664} (\bibinfo {year} {2013})}\BibitemShut {NoStop}%
\bibitem [{\citenamefont {Ni}\ \emph {et~al.}(2014)\citenamefont {Ni}, \citenamefont {Bortnik}, \citenamefont {Nishimura}, \citenamefont {Thorne}, \citenamefont {Li}, \citenamefont {Angelopoulos}, \citenamefont {Ebihara},\ and\ \citenamefont {Weatherwax}}]{ni_chorus_2014}%
  \BibitemOpen
  \bibfield  {author} {\bibinfo {author} {\bibfnamefont {B.}~\bibnamefont {Ni}}, \bibinfo {author} {\bibfnamefont {J.}~\bibnamefont {Bortnik}}, \bibinfo {author} {\bibfnamefont {Y.}~\bibnamefont {Nishimura}}, \bibinfo {author} {\bibfnamefont {R.~M.}\ \bibnamefont {Thorne}}, \bibinfo {author} {\bibfnamefont {W.}~\bibnamefont {Li}}, \bibinfo {author} {\bibfnamefont {V.}~\bibnamefont {Angelopoulos}}, \bibinfo {author} {\bibfnamefont {Y.}~\bibnamefont {Ebihara}},\ and\ \bibinfo {author} {\bibfnamefont {A.~T.}\ \bibnamefont {Weatherwax}},\ }\bibfield  {title} {\bibinfo {title} {Chorus wave scattering responsible for the {{Earth}}'s dayside diffuse auroral precipitation: {{A}} detailed case study},\ }\href {https://doi.org/10.1002/2013JA019507} {\bibfield  {journal} {\bibinfo  {journal} {Journal of Geophysical Research: Space Physics}\ }\textbf {\bibinfo {volume} {119}},\ \bibinfo {pages} {897} (\bibinfo {year} {2014})}\BibitemShut {NoStop}%
\bibitem [{\citenamefont {Ni}\ \emph {et~al.}(2008)\citenamefont {Ni}, \citenamefont {Thorne}, \citenamefont {Shprits},\ and\ \citenamefont {Bortnik}}]{ni_resonant_2008}%
  \BibitemOpen
  \bibfield  {author} {\bibinfo {author} {\bibfnamefont {B.}~\bibnamefont {Ni}}, \bibinfo {author} {\bibfnamefont {R.~M.}\ \bibnamefont {Thorne}}, \bibinfo {author} {\bibfnamefont {Y.~Y.}\ \bibnamefont {Shprits}},\ and\ \bibinfo {author} {\bibfnamefont {J.}~\bibnamefont {Bortnik}},\ }\bibfield  {title} {\bibinfo {title} {Resonant scattering of plasma sheet electrons by whistler-mode chorus: {{Contribution}} to diffuse auroral precipitation},\ }\bibfield  {journal} {\bibinfo  {journal} {Geophysical Research Letters}\ }\textbf {\bibinfo {volume} {35}},\ \href {https://doi.org/10.1029/2008GL034032} {10.1029/2008GL034032} (\bibinfo {year} {2008})\BibitemShut {NoStop}%
\bibitem [{\citenamefont {Ni}\ \emph {et~al.}(2011)\citenamefont {Ni}, \citenamefont {Thorne}, \citenamefont {Meredith}, \citenamefont {Horne},\ and\ \citenamefont {Shprits}}]{ni_resonant_2011}%
  \BibitemOpen
  \bibfield  {author} {\bibinfo {author} {\bibfnamefont {B.}~\bibnamefont {Ni}}, \bibinfo {author} {\bibfnamefont {R.~M.}\ \bibnamefont {Thorne}}, \bibinfo {author} {\bibfnamefont {N.~P.}\ \bibnamefont {Meredith}}, \bibinfo {author} {\bibfnamefont {R.~B.}\ \bibnamefont {Horne}},\ and\ \bibinfo {author} {\bibfnamefont {Y.~Y.}\ \bibnamefont {Shprits}},\ }\bibfield  {title} {\bibinfo {title} {Resonant scattering of plasma sheet electrons leading to diffuse auroral precipitation: 2. {{Evaluation}} for whistler mode chorus waves},\ }\bibfield  {journal} {\bibinfo  {journal} {Journal of Geophysical Research: Space Physics}\ }\textbf {\bibinfo {volume} {116}},\ \href {https://doi.org/10.1029/2010JA016233} {10.1029/2010JA016233} (\bibinfo {year} {2011})\BibitemShut {NoStop}%
\bibitem [{\citenamefont {Fang}\ \emph {et~al.}(2010)\citenamefont {Fang}, \citenamefont {Randall}, \citenamefont {Lummerzheim}, \citenamefont {Wang}, \citenamefont {Lu}, \citenamefont {Solomon},\ and\ \citenamefont {Frahm}}]{fangParameterizationMonoenergeticElectron2010}%
  \BibitemOpen
  \bibfield  {author} {\bibinfo {author} {\bibfnamefont {X.}~\bibnamefont {Fang}}, \bibinfo {author} {\bibfnamefont {C.~E.}\ \bibnamefont {Randall}}, \bibinfo {author} {\bibfnamefont {D.}~\bibnamefont {Lummerzheim}}, \bibinfo {author} {\bibfnamefont {W.}~\bibnamefont {Wang}}, \bibinfo {author} {\bibfnamefont {G.}~\bibnamefont {Lu}}, \bibinfo {author} {\bibfnamefont {S.~C.}\ \bibnamefont {Solomon}},\ and\ \bibinfo {author} {\bibfnamefont {R.~A.}\ \bibnamefont {Frahm}},\ }\bibfield  {title} {\bibinfo {title} {Parameterization of monoenergetic electron impact ionization},\ }\bibfield  {journal} {\bibinfo  {journal} {Geophysical Research Letters}\ }\textbf {\bibinfo {volume} {37}},\ \href {https://doi.org/10.1029/2010GL045406} {10.1029/2010GL045406} (\bibinfo {year} {2010})\BibitemShut {NoStop}%
\bibitem [{\citenamefont {Pr{\"o}lss}(2004)}]{prolssAbsorptionDissipationSolar2004}%
  \BibitemOpen
  \bibfield  {author} {\bibinfo {author} {\bibfnamefont {G.~W.}\ \bibnamefont {Pr{\"o}lss}},\ }\bibfield  {title} {\bibinfo {title} {Absorption and {{Dissipation}} of {{Solar Radiation Energy}}},\ }in\ \href {https://doi.org/10.1007/978-3-642-97123-5_3} {\emph {\bibinfo {booktitle} {Physics of the {{Earth}}'s {{Space Environment}}: {{An Introduction}}}}},\ \bibinfo {editor} {edited by\ \bibinfo {editor} {\bibfnamefont {G.~W.}\ \bibnamefont {Pr{\"o}lss}}}\ (\bibinfo  {publisher} {Springer},\ \bibinfo {address} {Berlin, Heidelberg},\ \bibinfo {year} {2004})\ pp.\ \bibinfo {pages} {77--157}\BibitemShut {NoStop}%
\bibitem [{\citenamefont {Richmond}\ and\ \citenamefont {Thayer}(2000)}]{richmondIonosphericElectrodynamicsTutorial2000}%
  \BibitemOpen
  \bibfield  {author} {\bibinfo {author} {\bibfnamefont {A.~D.}\ \bibnamefont {Richmond}}\ and\ \bibinfo {author} {\bibfnamefont {J.~P.}\ \bibnamefont {Thayer}},\ }\bibfield  {title} {\bibinfo {title} {Ionospheric {{Electrodynamics}}: {{A Tutorial}}},\ }in\ \href {https://doi.org/10.1029/GM118p0131} {\emph {\bibinfo {booktitle} {Magnetospheric {{Current Systems}}}}}\ (\bibinfo  {publisher} {American Geophysical Union (AGU)},\ \bibinfo {year} {2000})\ pp.\ \bibinfo {pages} {131--146}\BibitemShut {NoStop}%
\bibitem [{\citenamefont {Ivarsen}\ \emph {et~al.}(2025{\natexlab{a}})\citenamefont {Ivarsen}, \citenamefont {Huyghebaert}, \citenamefont {Jin}, \citenamefont {Miyashita}, \citenamefont {{St.-Maurice}}, \citenamefont {Hussey}, \citenamefont {Dan}, \citenamefont {Kasahara}, \citenamefont {Song}, \citenamefont {Jayachandran}, \citenamefont {Yokota}, \citenamefont {Miyoshi}, \citenamefont {Kasahara}, \citenamefont {Shinohara},\ and\ \citenamefont {Matsuoka}}]{ivarsen_transient_2025}%
  \BibitemOpen
  \bibfield  {author} {\bibinfo {author} {\bibfnamefont {M.}~\bibnamefont {Ivarsen}}, \bibinfo {author} {\bibfnamefont {D.}~\bibnamefont {Huyghebaert}}, \bibinfo {author} {\bibfnamefont {Y.}~\bibnamefont {Jin}}, \bibinfo {author} {\bibfnamefont {Y.}~\bibnamefont {Miyashita}}, \bibinfo {author} {\bibfnamefont {J.-P.}\ \bibnamefont {{St.-Maurice}}}, \bibinfo {author} {\bibfnamefont {G.}~\bibnamefont {Hussey}}, \bibinfo {author} {\bibfnamefont {B.}~\bibnamefont {Dan}}, \bibinfo {author} {\bibfnamefont {S.}~\bibnamefont {Kasahara}}, \bibinfo {author} {\bibfnamefont {K.}~\bibnamefont {Song}}, \bibinfo {author} {\bibfnamefont {P.}~\bibnamefont {Jayachandran}}, \bibinfo {author} {\bibfnamefont {S.}~\bibnamefont {Yokota}}, \bibinfo {author} {\bibfnamefont {Y.}~\bibnamefont {Miyoshi}}, \bibinfo {author} {\bibfnamefont {Y.}~\bibnamefont {Kasahara}}, \bibinfo {author} {\bibfnamefont {I.}~\bibnamefont {Shinohara}},\ and\ \bibinfo {author} {\bibfnamefont {A.}~\bibnamefont {Matsuoka}},\ }\href
  {https://doi.org/10.21203/rs.3.rs-5313766/v2} {\bibinfo {title} {Transient, {{Turbulent Hall Currents}} in the {{Sunlit Terrestrial Ionosphere}}}} (\bibinfo {year} {2025}{\natexlab{a}})\BibitemShut {NoStop}%
\bibitem [{\citenamefont {Hysell}(2015)}]{hysellRadarAurora2015}%
  \BibitemOpen
  \bibfield  {author} {\bibinfo {author} {\bibfnamefont {D.~L.}\ \bibnamefont {Hysell}},\ }\bibfield  {title} {\bibinfo {title} {The {{Radar Aurora}}},\ }in\ \href {https://doi.org/10.1002/9781118978719.ch14} {\emph {\bibinfo {booktitle} {Auroral {{Dynamics}} and {{Space Weather}}}}}\ (\bibinfo  {publisher} {American Geophysical Union (AGU)},\ \bibinfo {year} {2015})\ Chap.~\bibinfo {chapter} {14}, pp.\ \bibinfo {pages} {191--209}\BibitemShut {NoStop}%
\bibitem [{\citenamefont {Huyghebaert}\ \emph {et~al.}(2019)\citenamefont {Huyghebaert}, \citenamefont {Hussey}, \citenamefont {Vierinen}, \citenamefont {McWilliams},\ and\ \citenamefont {{St-Maurice}}}]{huyghebaertICEBEARAlldigitalBistatic2019}%
  \BibitemOpen
  \bibfield  {author} {\bibinfo {author} {\bibfnamefont {D.}~\bibnamefont {Huyghebaert}}, \bibinfo {author} {\bibfnamefont {G.}~\bibnamefont {Hussey}}, \bibinfo {author} {\bibfnamefont {J.}~\bibnamefont {Vierinen}}, \bibinfo {author} {\bibfnamefont {K.}~\bibnamefont {McWilliams}},\ and\ \bibinfo {author} {\bibfnamefont {J.-P.}\ \bibnamefont {{St-Maurice}}},\ }\bibfield  {title} {\bibinfo {title} {{{ICEBEAR}}: {{An}} all-digital bistatic coded continuous-wave radar for studies of the {{E}} region of the ionosphere},\ }\href {https://doi.org/10.1029/2018RS006747} {\bibfield  {journal} {\bibinfo  {journal} {Radio Science}\ }\textbf {\bibinfo {volume} {54}},\ \bibinfo {pages} {349} (\bibinfo {year} {2019})}\BibitemShut {NoStop}%
\bibitem [{\citenamefont {Farley}(1963)}]{farleyPlasmaInstabilityResulting1963}%
  \BibitemOpen
  \bibfield  {author} {\bibinfo {author} {\bibfnamefont {D.~T.}\ \bibnamefont {Farley}},\ }\bibfield  {title} {\bibinfo {title} {A plasma instability resulting in field-aligned irregularities in the ionosphere},\ }\href {https://doi.org/10.1029/JZ068i022p06083} {\bibfield  {journal} {\bibinfo  {journal} {Journal of Geophysical Research (1896-1977)}\ }\textbf {\bibinfo {volume} {68}},\ \bibinfo {pages} {6083} (\bibinfo {year} {1963})}\BibitemShut {NoStop}%
\bibitem [{\citenamefont {Buneman}(1963)}]{bunemanExcitationFieldAligned1963}%
  \BibitemOpen
  \bibfield  {author} {\bibinfo {author} {\bibfnamefont {O.}~\bibnamefont {Buneman}},\ }\bibfield  {title} {\bibinfo {title} {Excitation of {{Field Aligned Sound Waves}} by {{Electron Streams}}},\ }\href {https://doi.org/10.1103/PhysRevLett.10.285} {\bibfield  {journal} {\bibinfo  {journal} {Physical Review Letters}\ }\textbf {\bibinfo {volume} {10}},\ \bibinfo {pages} {285} (\bibinfo {year} {1963})}\BibitemShut {NoStop}%
\bibitem [{\citenamefont {Prikryl}\ \emph {et~al.}(1988)\citenamefont {Prikryl}, \citenamefont {Andr{\'e}}, \citenamefont {Sofko},\ and\ \citenamefont {Koehler}}]{prikryl_doppler_1988}%
  \BibitemOpen
  \bibfield  {author} {\bibinfo {author} {\bibfnamefont {P.}~\bibnamefont {Prikryl}}, \bibinfo {author} {\bibfnamefont {D.}~\bibnamefont {Andr{\'e}}}, \bibinfo {author} {\bibfnamefont {G.~J.}\ \bibnamefont {Sofko}},\ and\ \bibinfo {author} {\bibfnamefont {J.~A.}\ \bibnamefont {Koehler}},\ }\bibfield  {title} {\bibinfo {title} {Doppler radar observations of harmonics of electrostatic ion cyclotron waves in the auroral ionosphere},\ }\href {https://doi.org/10.1029/JA093iA07p07409} {\bibfield  {journal} {\bibinfo  {journal} {Journal of Geophysical Research: Space Physics}\ }\textbf {\bibinfo {volume} {93}},\ \bibinfo {pages} {7409} (\bibinfo {year} {1988})}\BibitemShut {NoStop}%
\bibitem [{\citenamefont {Prikryl}\ \emph {et~al.}(1990)\citenamefont {Prikryl}, \citenamefont {Andre}, \citenamefont {Koehler}, \citenamefont {Sofko},\ and\ \citenamefont {McKibben}}]{prikryl_evidence_1990}%
  \BibitemOpen
  \bibfield  {author} {\bibinfo {author} {\bibfnamefont {P.}~\bibnamefont {Prikryl}}, \bibinfo {author} {\bibfnamefont {D.}~\bibnamefont {Andre}}, \bibinfo {author} {\bibfnamefont {J.~A.}\ \bibnamefont {Koehler}}, \bibinfo {author} {\bibfnamefont {G.~J.}\ \bibnamefont {Sofko}},\ and\ \bibinfo {author} {\bibfnamefont {M.~J.}\ \bibnamefont {McKibben}},\ }\bibfield  {title} {\bibinfo {title} {Evidence of highly localized auroral scatterers from 50-{{MHz CW}} radar interferometry},\ }\href@noop {} {\bibfield  {journal} {\bibinfo  {journal} {Planetary and space science}\ }\textbf {\bibinfo {volume} {38}},\ \bibinfo {pages} {933} (\bibinfo {year} {1990})}\BibitemShut {NoStop}%
\bibitem [{\citenamefont {Ivarsen}\ \emph {et~al.}(2024{\natexlab{a}})\citenamefont {Ivarsen}, \citenamefont {{St-Maurice}}, \citenamefont {Hussey}, \citenamefont {Huyghebaert},\ and\ \citenamefont {Gillies}}]{ivarsen_point-cloud_2024}%
  \BibitemOpen
  \bibfield  {author} {\bibinfo {author} {\bibfnamefont {M.~F.}\ \bibnamefont {Ivarsen}}, \bibinfo {author} {\bibfnamefont {J.-P.}\ \bibnamefont {{St-Maurice}}}, \bibinfo {author} {\bibfnamefont {G.~C.}\ \bibnamefont {Hussey}}, \bibinfo {author} {\bibfnamefont {D.~R.}\ \bibnamefont {Huyghebaert}},\ and\ \bibinfo {author} {\bibfnamefont {M.~D.}\ \bibnamefont {Gillies}},\ }\bibfield  {title} {\bibinfo {title} {Point-cloud clustering and tracking algorithm for radar interferometry},\ }\href {https://doi.org/10.1103/PhysRevE.110.045207} {\bibfield  {journal} {\bibinfo  {journal} {Physical Review E}\ }\textbf {\bibinfo {volume} {110}},\ \bibinfo {pages} {045207} (\bibinfo {year} {2024}{\natexlab{a}})}\BibitemShut {NoStop}%
\bibitem [{\citenamefont {Ivarsen}\ \emph {et~al.}(2024{\natexlab{b}})\citenamefont {Ivarsen}, \citenamefont {{St-Maurice}}, \citenamefont {Huyghebaert}, \citenamefont {Gillies}, \citenamefont {Lind}, \citenamefont {Pitzel},\ and\ \citenamefont {Hussey}}]{ivarsen_deriving_2024}%
  \BibitemOpen
  \bibfield  {author} {\bibinfo {author} {\bibfnamefont {M.~F.}\ \bibnamefont {Ivarsen}}, \bibinfo {author} {\bibfnamefont {J.-P.}\ \bibnamefont {{St-Maurice}}}, \bibinfo {author} {\bibfnamefont {D.~R.}\ \bibnamefont {Huyghebaert}}, \bibinfo {author} {\bibfnamefont {M.~D.}\ \bibnamefont {Gillies}}, \bibinfo {author} {\bibfnamefont {F.}~\bibnamefont {Lind}}, \bibinfo {author} {\bibfnamefont {B.}~\bibnamefont {Pitzel}},\ and\ \bibinfo {author} {\bibfnamefont {G.~C.}\ \bibnamefont {Hussey}},\ }\bibfield  {title} {\bibinfo {title} {Deriving the {{Ionospheric Electric Field From}} the {{Bulk Motion}} of {{Radar Aurora}} in the {{E-Region}}},\ }\href {https://doi.org/10.1029/2024JA033060} {\bibfield  {journal} {\bibinfo  {journal} {Journal of Geophysical Research: Space Physics}\ }\textbf {\bibinfo {volume} {129}},\ \bibinfo {pages} {e2024JA033060} (\bibinfo {year} {2024}{\natexlab{b}})}\BibitemShut {NoStop}%
\bibitem [{\citenamefont {Papitashvili}\ and\ \citenamefont {King}(2020)}]{papitashviliOMNIHourlyData2020}%
  \BibitemOpen
  \bibfield  {author} {\bibinfo {author} {\bibfnamefont {N.~E.}\ \bibnamefont {Papitashvili}}\ and\ \bibinfo {author} {\bibfnamefont {J.~H.}\ \bibnamefont {King}},\ }\href {https://doi.org/10.48322/1SHR-HT18} {\bibinfo {title} {{{OMNI Hourly Data Set}}}} (\bibinfo {year} {2020})\BibitemShut {NoStop}%
\bibitem [{\citenamefont {Davis}(2007)}]{davis_history_2007}%
  \BibitemOpen
  \bibfield  {author} {\bibinfo {author} {\bibfnamefont {G.~K.}\ \bibnamefont {Davis}},\ }\bibfield  {title} {\bibinfo {title} {History of the {{NOAA}} satellite program},\ }\href {https://doi.org/10.1117/1.2642347} {\bibfield  {journal} {\bibinfo  {journal} {Journal of Applied Remote Sensing}\ }\textbf {\bibinfo {volume} {1}},\ \bibinfo {pages} {012504} (\bibinfo {year} {2007})}\BibitemShut {NoStop}%
\bibitem [{\citenamefont {Evans}(2000)}]{evans_polar_2000}%
  \BibitemOpen
  \bibfield  {author} {\bibinfo {author} {\bibfnamefont {D.~S. D.~S.}\ \bibnamefont {Evans}},\ }\bibfield  {title} {\bibinfo {title} {Polar orbiting environmental satellite space environment monitor-2 : Instrument description and archive data},\ }\href@noop {} {\bibfield  {journal} {\bibinfo  {journal} {NOAA technical memorandum OAR SEC ; 93}\ } (\bibinfo {year} {2000})}\BibitemShut {NoStop}%
\bibitem [{\citenamefont {Baker}\ and\ \citenamefont {Wing}(1989)}]{bakerNewMagneticCoordinate1989}%
  \BibitemOpen
  \bibfield  {author} {\bibinfo {author} {\bibfnamefont {K.~B.}\ \bibnamefont {Baker}}\ and\ \bibinfo {author} {\bibfnamefont {S.}~\bibnamefont {Wing}},\ }\bibfield  {title} {\bibinfo {title} {A new magnetic coordinate system for conjugate studies at high latitudes},\ }\href {https://doi.org/10.1029/JA094iA07p09139} {\bibfield  {journal} {\bibinfo  {journal} {Journal of Geophysical Research: Space Physics}\ }\textbf {\bibinfo {volume} {94}},\ \bibinfo {pages} {9139} (\bibinfo {year} {1989})}\BibitemShut {NoStop}%
\bibitem [{\citenamefont {Li}\ \emph {et~al.}(2024)\citenamefont {Li}, \citenamefont {Zhou}, \citenamefont {Wang}, \citenamefont {Liu}, \citenamefont {Zong}, \citenamefont {Yao}, \citenamefont {Artemyev}, \citenamefont {Omura}, \citenamefont {Li}, \citenamefont {Yue},\ and\ \citenamefont {Shi}}]{li_bow_2024}%
  \BibitemOpen
  \bibfield  {author} {\bibinfo {author} {\bibfnamefont {J.-H.}\ \bibnamefont {Li}}, \bibinfo {author} {\bibfnamefont {X.-Z.}\ \bibnamefont {Zhou}}, \bibinfo {author} {\bibfnamefont {S.}~\bibnamefont {Wang}}, \bibinfo {author} {\bibfnamefont {Z.-Y.}\ \bibnamefont {Liu}}, \bibinfo {author} {\bibfnamefont {Q.-G.}\ \bibnamefont {Zong}}, \bibinfo {author} {\bibfnamefont {S.-T.}\ \bibnamefont {Yao}}, \bibinfo {author} {\bibfnamefont {A.~V.}\ \bibnamefont {Artemyev}}, \bibinfo {author} {\bibfnamefont {Y.}~\bibnamefont {Omura}}, \bibinfo {author} {\bibfnamefont {L.}~\bibnamefont {Li}}, \bibinfo {author} {\bibfnamefont {C.}~\bibnamefont {Yue}},\ and\ \bibinfo {author} {\bibfnamefont {Q.-Q.}\ \bibnamefont {Shi}},\ }\bibfield  {title} {\bibinfo {title} {Bow {{Shock Ripples}} and {{Their Modulation}} of {{Whistler Wave Packets}}: {{MMS Observations}}},\ }\href {https://doi.org/10.1029/2024GL111590} {\bibfield  {journal} {\bibinfo  {journal} {Geophysical Research Letters}\ }\textbf {\bibinfo {volume} {51}},\ \bibinfo
  {pages} {e2024GL111590} (\bibinfo {year} {2024})}\BibitemShut {NoStop}%
\bibitem [{\citenamefont {Kr{\"a}mer}\ \emph {et~al.}(2024)\citenamefont {Kr{\"a}mer}, \citenamefont {Koller}, \citenamefont {Suni}, \citenamefont {LaMoury}, \citenamefont {P{\"o}ppelwerth}, \citenamefont {Glebe}, \citenamefont {{Mohammed-Amin}}, \citenamefont {Raptis}, \citenamefont {Vuorinen}, \citenamefont {Weiss}, \citenamefont {Xirogiannopoulou}, \citenamefont {Archer}, \citenamefont {{Blanco-Cano}}, \citenamefont {Gunell}, \citenamefont {Hietala}, \citenamefont {Karlsson}, \citenamefont {Plaschke}, \citenamefont {Preisser}, \citenamefont {Roberts}, \citenamefont {Simon~Wedlund}, \citenamefont {Temmer},\ and\ \citenamefont {V{\"o}r{\"o}s}}]{kramer_jets_2024}%
  \BibitemOpen
  \bibfield  {author} {\bibinfo {author} {\bibfnamefont {E.}~\bibnamefont {Kr{\"a}mer}}, \bibinfo {author} {\bibfnamefont {F.}~\bibnamefont {Koller}}, \bibinfo {author} {\bibfnamefont {J.}~\bibnamefont {Suni}}, \bibinfo {author} {\bibfnamefont {A.~T.}\ \bibnamefont {LaMoury}}, \bibinfo {author} {\bibfnamefont {A.}~\bibnamefont {P{\"o}ppelwerth}}, \bibinfo {author} {\bibfnamefont {G.}~\bibnamefont {Glebe}}, \bibinfo {author} {\bibfnamefont {T.}~\bibnamefont {{Mohammed-Amin}}}, \bibinfo {author} {\bibfnamefont {S.}~\bibnamefont {Raptis}}, \bibinfo {author} {\bibfnamefont {L.}~\bibnamefont {Vuorinen}}, \bibinfo {author} {\bibfnamefont {S.}~\bibnamefont {Weiss}}, \bibinfo {author} {\bibfnamefont {N.}~\bibnamefont {Xirogiannopoulou}}, \bibinfo {author} {\bibfnamefont {M.}~\bibnamefont {Archer}}, \bibinfo {author} {\bibfnamefont {X.}~\bibnamefont {{Blanco-Cano}}}, \bibinfo {author} {\bibfnamefont {H.}~\bibnamefont {Gunell}}, \bibinfo {author} {\bibfnamefont {H.}~\bibnamefont {Hietala}}, \bibinfo {author}
  {\bibfnamefont {T.}~\bibnamefont {Karlsson}}, \bibinfo {author} {\bibfnamefont {F.}~\bibnamefont {Plaschke}}, \bibinfo {author} {\bibfnamefont {L.}~\bibnamefont {Preisser}}, \bibinfo {author} {\bibfnamefont {O.}~\bibnamefont {Roberts}}, \bibinfo {author} {\bibfnamefont {C.}~\bibnamefont {Simon~Wedlund}}, \bibinfo {author} {\bibfnamefont {M.}~\bibnamefont {Temmer}},\ and\ \bibinfo {author} {\bibfnamefont {Z.}~\bibnamefont {V{\"o}r{\"o}s}},\ }\bibfield  {title} {\bibinfo {title} {Jets {{Downstream}} of {{Collisionless Shocks}}: {{Recent Discoveries}} and {{Challenges}}},\ }\href {https://doi.org/10.1007/s11214-024-01129-3} {\bibfield  {journal} {\bibinfo  {journal} {Space Science Reviews}\ }\textbf {\bibinfo {volume} {221}},\ \bibinfo {pages} {4} (\bibinfo {year} {2024})}\BibitemShut {NoStop}%
\bibitem [{\citenamefont {Newell}\ and\ \citenamefont {Meng}(1988)}]{newellCuspCleftBoundary1988}%
  \BibitemOpen
  \bibfield  {author} {\bibinfo {author} {\bibfnamefont {P.~T.}\ \bibnamefont {Newell}}\ and\ \bibinfo {author} {\bibfnamefont {C.-I.}\ \bibnamefont {Meng}},\ }\bibfield  {title} {\bibinfo {title} {The cusp and the cleft/boundary layer: {{Low-altitude}} identification and statistical local time variation},\ }\href {https://doi.org/10.1029/JA093iA12p14549} {\bibfield  {journal} {\bibinfo  {journal} {Journal of Geophysical Research: Space Physics}\ }\textbf {\bibinfo {volume} {93}},\ \bibinfo {pages} {14549} (\bibinfo {year} {1988})}\BibitemShut {NoStop}%
\bibitem [{\citenamefont {Ivarsen}\ \emph {et~al.}(2023)\citenamefont {Ivarsen}, \citenamefont {Jin}, \citenamefont {Spicher}, \citenamefont {{St-Maurice}}, \citenamefont {Park},\ and\ \citenamefont {Billett}}]{ivarsenGNSSScintillationsCusp2023}%
  \BibitemOpen
  \bibfield  {author} {\bibinfo {author} {\bibfnamefont {M.~F.}\ \bibnamefont {Ivarsen}}, \bibinfo {author} {\bibfnamefont {Y.}~\bibnamefont {Jin}}, \bibinfo {author} {\bibfnamefont {A.}~\bibnamefont {Spicher}}, \bibinfo {author} {\bibfnamefont {J.-P.}\ \bibnamefont {{St-Maurice}}}, \bibinfo {author} {\bibfnamefont {J.}~\bibnamefont {Park}},\ and\ \bibinfo {author} {\bibfnamefont {D.}~\bibnamefont {Billett}},\ }\bibfield  {title} {\bibinfo {title} {{{GNSS Scintillations}} in the {{Cusp}}, and the {{Role}} of {{Precipitating Particle Energy Fluxes}}},\ }\href {https://doi.org/10.1029/2023JA031849} {\bibfield  {journal} {\bibinfo  {journal} {Journal of Geophysical Research: Space Physics}\ }\textbf {\bibinfo {volume} {128}},\ \bibinfo {pages} {e2023JA031849} (\bibinfo {year} {2023})}\BibitemShut {NoStop}%
\bibitem [{\citenamefont {Asai}\ \emph {et~al.}(2005)\citenamefont {Asai}, \citenamefont {Maezawa}, \citenamefont {Mukai},\ and\ \citenamefont {Hayakawa}}]{asai_latitudinal_2005}%
  \BibitemOpen
  \bibfield  {author} {\bibinfo {author} {\bibfnamefont {K.~T.}\ \bibnamefont {Asai}}, \bibinfo {author} {\bibfnamefont {K.}~\bibnamefont {Maezawa}}, \bibinfo {author} {\bibfnamefont {T.}~\bibnamefont {Mukai}},\ and\ \bibinfo {author} {\bibfnamefont {H.}~\bibnamefont {Hayakawa}},\ }\bibfield  {title} {\bibinfo {title} {Latitudinal and longitudinal displacement of cusp ion precipitation controlled by {{IMF By}} and {{Bz}}},\ }\href {https://doi.org/10.1186/BF03351842} {\bibfield  {journal} {\bibinfo  {journal} {Earth, Planets and Space}\ }\textbf {\bibinfo {volume} {57}},\ \bibinfo {pages} {627} (\bibinfo {year} {2005})}\BibitemShut {NoStop}%
\bibitem [{\citenamefont {Jin}\ \emph {et~al.}(2017)\citenamefont {Jin}, \citenamefont {Moen}, \citenamefont {Oksavik}, \citenamefont {Spicher}, \citenamefont {Clausen},\ and\ \citenamefont {Miloch}}]{jinGPSScintillationsAssociated2017}%
  \BibitemOpen
  \bibfield  {author} {\bibinfo {author} {\bibfnamefont {Y.}~\bibnamefont {Jin}}, \bibinfo {author} {\bibfnamefont {J.~I.}\ \bibnamefont {Moen}}, \bibinfo {author} {\bibfnamefont {K.}~\bibnamefont {Oksavik}}, \bibinfo {author} {\bibfnamefont {A.}~\bibnamefont {Spicher}}, \bibinfo {author} {\bibfnamefont {L.~B.~N.}\ \bibnamefont {Clausen}},\ and\ \bibinfo {author} {\bibfnamefont {W.~J.}\ \bibnamefont {Miloch}},\ }\bibfield  {title} {\bibinfo {title} {{{GPS}} scintillations associated with cusp dynamics and polar cap patches},\ }\href {https://doi.org/10.1051/swsc/2017022} {\bibfield  {journal} {\bibinfo  {journal} {Journal of Space Weather and Space Climate}\ }\textbf {\bibinfo {volume} {7}},\ \bibinfo {pages} {A23} (\bibinfo {year} {2017})}\BibitemShut {NoStop}%
\bibitem [{\citenamefont {Nielsen}\ and\ \citenamefont {Schlegel}(1983)}]{nielsen_first_1983}%
  \BibitemOpen
  \bibfield  {author} {\bibinfo {author} {\bibfnamefont {E.}~\bibnamefont {Nielsen}}\ and\ \bibinfo {author} {\bibfnamefont {K.}~\bibnamefont {Schlegel}},\ }\bibfield  {title} {\bibinfo {title} {A first comparison of {{STARE}} and {{EISCAT}} electron drift velocity measurements},\ }\href {https://doi.org/10.1029/JA088iA07p05745} {\bibfield  {journal} {\bibinfo  {journal} {Journal of Geophysical Research}\ }\textbf {\bibinfo {volume} {88}},\ \bibinfo {pages} {5745} (\bibinfo {year} {1983})}\BibitemShut {NoStop}%
\bibitem [{\citenamefont {Foster}\ and\ \citenamefont {Erickson}(2000)}]{fosterSimultaneousObservationsEregion2000}%
  \BibitemOpen
  \bibfield  {author} {\bibinfo {author} {\bibfnamefont {J.~C.}\ \bibnamefont {Foster}}\ and\ \bibinfo {author} {\bibfnamefont {P.~J.}\ \bibnamefont {Erickson}},\ }\bibfield  {title} {\bibinfo {title} {Simultaneous observations of {{E-region}} coherent backscatter and electric field amplitude at {{F-region}} heights with the {{Millstone Hill UHF Radar}}},\ }\href {https://doi.org/10.1029/2000GL000042} {\bibfield  {journal} {\bibinfo  {journal} {Geophysical Research Letters}\ }\textbf {\bibinfo {volume} {27}},\ \bibinfo {pages} {3177} (\bibinfo {year} {2000})}\BibitemShut {NoStop}%
\bibitem [{\citenamefont {Archer}\ \emph {et~al.}(2019)\citenamefont {Archer}, \citenamefont {Hietala}, \citenamefont {Hartinger}, \citenamefont {Plaschke},\ and\ \citenamefont {Angelopoulos}}]{archer_direct_2019}%
  \BibitemOpen
  \bibfield  {author} {\bibinfo {author} {\bibfnamefont {M.~O.}\ \bibnamefont {Archer}}, \bibinfo {author} {\bibfnamefont {H.}~\bibnamefont {Hietala}}, \bibinfo {author} {\bibfnamefont {M.~D.}\ \bibnamefont {Hartinger}}, \bibinfo {author} {\bibfnamefont {F.}~\bibnamefont {Plaschke}},\ and\ \bibinfo {author} {\bibfnamefont {V.}~\bibnamefont {Angelopoulos}},\ }\bibfield  {title} {\bibinfo {title} {Direct observations of a surface eigenmode of the dayside magnetopause},\ }\href {https://doi.org/10.1038/s41467-018-08134-5} {\bibfield  {journal} {\bibinfo  {journal} {Nature Communications}\ }\textbf {\bibinfo {volume} {10}},\ \bibinfo {pages} {615} (\bibinfo {year} {2019})}\BibitemShut {NoStop}%
\bibitem [{\citenamefont {Qiu}\ \emph {et~al.}(2024)\citenamefont {Qiu}, \citenamefont {Han}, \citenamefont {Shi},\ and\ \citenamefont {Liu}}]{qiu_magnetosheath_2024}%
  \BibitemOpen
  \bibfield  {author} {\bibinfo {author} {\bibfnamefont {H.-X.}\ \bibnamefont {Qiu}}, \bibinfo {author} {\bibfnamefont {D.-S.}\ \bibnamefont {Han}}, \bibinfo {author} {\bibfnamefont {R.}~\bibnamefont {Shi}},\ and\ \bibinfo {author} {\bibfnamefont {J.}~\bibnamefont {Liu}},\ }\bibfield  {title} {\bibinfo {title} {Magnetosheath {{High-Speed Jet Drives Multiple Auroral Arcs Near Local Noon}}},\ }\href {https://doi.org/10.1029/2024AV001197} {\bibfield  {journal} {\bibinfo  {journal} {AGU Advances}\ }\textbf {\bibinfo {volume} {5}},\ \bibinfo {pages} {e2024AV001197} (\bibinfo {year} {2024})}\BibitemShut {NoStop}%
\bibitem [{\citenamefont {Shi}\ \emph {et~al.}(2020)\citenamefont {Shi}, \citenamefont {Shen}, \citenamefont {Tian}, \citenamefont {Degeling}, \citenamefont {Zong}, \citenamefont {Fu}, \citenamefont {Pu}, \citenamefont {Zhao}, \citenamefont {Zhang},\ and\ \citenamefont {Yao}}]{shi_magnetosphere_2020}%
  \BibitemOpen
  \bibfield  {author} {\bibinfo {author} {\bibfnamefont {Q.~Q.}\ \bibnamefont {Shi}}, \bibinfo {author} {\bibfnamefont {X.-C.}\ \bibnamefont {Shen}}, \bibinfo {author} {\bibfnamefont {A.~M.}\ \bibnamefont {Tian}}, \bibinfo {author} {\bibfnamefont {A.~W.}\ \bibnamefont {Degeling}}, \bibinfo {author} {\bibfnamefont {Q.}~\bibnamefont {Zong}}, \bibinfo {author} {\bibfnamefont {S.~Y.}\ \bibnamefont {Fu}}, \bibinfo {author} {\bibfnamefont {Z.~Y.}\ \bibnamefont {Pu}}, \bibinfo {author} {\bibfnamefont {H.~Y.}\ \bibnamefont {Zhao}}, \bibinfo {author} {\bibfnamefont {H.}~\bibnamefont {Zhang}},\ and\ \bibinfo {author} {\bibfnamefont {S.~T.}\ \bibnamefont {Yao}},\ }\bibfield  {title} {\bibinfo {title} {Magnetosphere {{Response}} to {{Solar Wind Dynamic Pressure Change}}},\ }in\ \href {https://doi.org/10.1002/9781119509592.ch5} {\emph {\bibinfo {booktitle} {Dayside {{Magnetosphere Interactions}}}}}\ (\bibinfo  {publisher} {American Geophysical Union (AGU)},\ \bibinfo {year} {2020})\ Chap.~\bibinfo {chapter} {5}, pp.\
  \bibinfo {pages} {77--97}\BibitemShut {NoStop}%
\bibitem [{\citenamefont {Lanchester}\ \emph {et~al.}(1996)\citenamefont {Lanchester}, \citenamefont {Kaila},\ and\ \citenamefont {McCrea}}]{lanchester_relationship_1996}%
  \BibitemOpen
  \bibfield  {author} {\bibinfo {author} {\bibfnamefont {B.~S.}\ \bibnamefont {Lanchester}}, \bibinfo {author} {\bibfnamefont {K.}~\bibnamefont {Kaila}},\ and\ \bibinfo {author} {\bibfnamefont {I.~W.}\ \bibnamefont {McCrea}},\ }\bibfield  {title} {\bibinfo {title} {Relationship between large horizontal electric fields and auroral arc elements},\ }\href {https://doi.org/10.1029/95JA02055} {\bibfield  {journal} {\bibinfo  {journal} {Journal of Geophysical Research: Space Physics}\ }\textbf {\bibinfo {volume} {101}},\ \bibinfo {pages} {5075} (\bibinfo {year} {1996})}\BibitemShut {NoStop}%
\bibitem [{\citenamefont {Opgenoorth}\ \emph {et~al.}(1990)\citenamefont {Opgenoorth}, \citenamefont {H{\"a}gstr{\"o}m}, \citenamefont {Williams},\ and\ \citenamefont {Jones}}]{opgenoorthRegionsStronglyEnhanced1990}%
  \BibitemOpen
  \bibfield  {author} {\bibinfo {author} {\bibfnamefont {H.~J.}\ \bibnamefont {Opgenoorth}}, \bibinfo {author} {\bibfnamefont {I.}~\bibnamefont {H{\"a}gstr{\"o}m}}, \bibinfo {author} {\bibfnamefont {P.~J.~S.}\ \bibnamefont {Williams}},\ and\ \bibinfo {author} {\bibfnamefont {G.~O.~L.}\ \bibnamefont {Jones}},\ }\bibfield  {title} {\bibinfo {title} {Regions of strongly enhanced perpendicular electric fields adjacent to auroral arcs},\ }\href {https://doi.org/10.1016/0021-9169(90)90044-N} {\bibfield  {journal} {\bibinfo  {journal} {Journal of Atmospheric and Terrestrial Physics}\ }\bibinfo {series} {The {{Fourth International EISCAT Workshop}}},\ \textbf {\bibinfo {volume} {52}},\ \bibinfo {pages} {449} (\bibinfo {year} {1990})}\BibitemShut {NoStop}%
\bibitem [{\citenamefont {Tuttle}\ \emph {et~al.}(2020)\citenamefont {Tuttle}, \citenamefont {Lanchester}, \citenamefont {Gustavsson}, \citenamefont {Whiter}, \citenamefont {Ivchenko}, \citenamefont {Fear},\ and\ \citenamefont {Lester}}]{tuttle_horizontal_2020}%
  \BibitemOpen
  \bibfield  {author} {\bibinfo {author} {\bibfnamefont {S.}~\bibnamefont {Tuttle}}, \bibinfo {author} {\bibfnamefont {B.}~\bibnamefont {Lanchester}}, \bibinfo {author} {\bibfnamefont {B.}~\bibnamefont {Gustavsson}}, \bibinfo {author} {\bibfnamefont {D.}~\bibnamefont {Whiter}}, \bibinfo {author} {\bibfnamefont {N.}~\bibnamefont {Ivchenko}}, \bibinfo {author} {\bibfnamefont {R.}~\bibnamefont {Fear}},\ and\ \bibinfo {author} {\bibfnamefont {M.}~\bibnamefont {Lester}},\ }\bibfield  {title} {\bibinfo {title} {Horizontal electric fields from flow of auroral {{O}}{\textsuperscript{+}}({\textsuperscript{2}}{{P}}) ions at sub-second temporal resolution},\ }\href {https://doi.org/10.5194/angeo-38-845-2020} {\bibfield  {journal} {\bibinfo  {journal} {Annales Geophysicae}\ }\textbf {\bibinfo {volume} {38}},\ \bibinfo {pages} {845} (\bibinfo {year} {2020})}\BibitemShut {NoStop}%
\bibitem [{\citenamefont {Krcelic}\ \emph {et~al.}(2024)\citenamefont {Krcelic}, \citenamefont {Fear}, \citenamefont {Whiter}, \citenamefont {Lanchester},\ and\ \citenamefont {Brindley}}]{krcelic_variability_2024}%
  \BibitemOpen
  \bibfield  {author} {\bibinfo {author} {\bibfnamefont {P.}~\bibnamefont {Krcelic}}, \bibinfo {author} {\bibfnamefont {R.~C.}\ \bibnamefont {Fear}}, \bibinfo {author} {\bibfnamefont {D.}~\bibnamefont {Whiter}}, \bibinfo {author} {\bibfnamefont {B.}~\bibnamefont {Lanchester}},\ and\ \bibinfo {author} {\bibfnamefont {N.}~\bibnamefont {Brindley}},\ }\bibfield  {title} {\bibinfo {title} {Variability in the {{Electrodynamics}} of the {{Small Scale Auroral Arc}}},\ }\href {https://doi.org/10.1029/2024JA032623} {\bibfield  {journal} {\bibinfo  {journal} {Journal of Geophysical Research: Space Physics}\ }\textbf {\bibinfo {volume} {129}},\ \bibinfo {pages} {e2024JA032623} (\bibinfo {year} {2024})}\BibitemShut {NoStop}%
\bibitem [{\citenamefont {Billett}\ \emph {et~al.}(2022)\citenamefont {Billett}, \citenamefont {McWilliams}, \citenamefont {Pakhotin}, \citenamefont {Burchill}, \citenamefont {Knudsen},\ and\ \citenamefont {Martin}}]{billettHighResolutionPoyntingFlux2022}%
  \BibitemOpen
  \bibfield  {author} {\bibinfo {author} {\bibfnamefont {D.~D.}\ \bibnamefont {Billett}}, \bibinfo {author} {\bibfnamefont {K.~A.}\ \bibnamefont {McWilliams}}, \bibinfo {author} {\bibfnamefont {I.~P.}\ \bibnamefont {Pakhotin}}, \bibinfo {author} {\bibfnamefont {J.~K.}\ \bibnamefont {Burchill}}, \bibinfo {author} {\bibfnamefont {D.~J.}\ \bibnamefont {Knudsen}},\ and\ \bibinfo {author} {\bibfnamefont {C.~J.}\ \bibnamefont {Martin}},\ }\bibfield  {title} {\bibinfo {title} {High-{{Resolution Poynting Flux Statistics From}} the {{Swarm Mission}}: {{How Much Is Being Underestimated}} at {{Larger Scales}}?},\ }\href {https://doi.org/10.1029/2022JA030573} {\bibfield  {journal} {\bibinfo  {journal} {Journal of Geophysical Research: Space Physics}\ }\textbf {\bibinfo {volume} {127}},\ \bibinfo {pages} {e2022JA030573} (\bibinfo {year} {2022})}\BibitemShut {NoStop}%
\bibitem [{\citenamefont {Liou}\ \emph {et~al.}(2001)\citenamefont {Liou}, \citenamefont {Newell},\ and\ \citenamefont {Meng}}]{liouSeasonalEffectsAuroral2001}%
  \BibitemOpen
  \bibfield  {author} {\bibinfo {author} {\bibfnamefont {K.}~\bibnamefont {Liou}}, \bibinfo {author} {\bibfnamefont {P.~T.}\ \bibnamefont {Newell}},\ and\ \bibinfo {author} {\bibfnamefont {C.-I.}\ \bibnamefont {Meng}},\ }\bibfield  {title} {\bibinfo {title} {Seasonal effects on auroral particle acceleration and precipitation},\ }\href {https://doi.org/10.1029/1999JA000391} {\bibfield  {journal} {\bibinfo  {journal} {Journal of Geophysical Research: Space Physics}\ }\textbf {\bibinfo {volume} {106}},\ \bibinfo {pages} {5531} (\bibinfo {year} {2001})}\BibitemShut {NoStop}%
\bibitem [{\citenamefont {Senior}\ \emph {et~al.}(1982)\citenamefont {Senior}, \citenamefont {Robinson},\ and\ \citenamefont {Potemra}}]{seniorRelationshipFieldalignedCurrents1982}%
  \BibitemOpen
  \bibfield  {author} {\bibinfo {author} {\bibfnamefont {C.}~\bibnamefont {Senior}}, \bibinfo {author} {\bibfnamefont {R.~M.}\ \bibnamefont {Robinson}},\ and\ \bibinfo {author} {\bibfnamefont {T.~A.}\ \bibnamefont {Potemra}},\ }\bibfield  {title} {\bibinfo {title} {Relationship between field-aligned currents, diffuse auroral precipitation and the westward electrojet in the early morning sector},\ }\href {https://doi.org/10.1029/JA087iA12p10469} {\bibfield  {journal} {\bibinfo  {journal} {Journal of Geophysical Research: Space Physics}\ }\textbf {\bibinfo {volume} {87}},\ \bibinfo {pages} {10469} (\bibinfo {year} {1982})}\BibitemShut {NoStop}%
\bibitem [{\citenamefont {Hosokawa}\ \emph {et~al.}(2010)\citenamefont {Hosokawa}, \citenamefont {Ogawa}, \citenamefont {Kadokura}, \citenamefont {Miyaoka},\ and\ \citenamefont {Sato}}]{hosokawaModulationIonosphericConductance2010}%
  \BibitemOpen
  \bibfield  {author} {\bibinfo {author} {\bibfnamefont {K.}~\bibnamefont {Hosokawa}}, \bibinfo {author} {\bibfnamefont {Y.}~\bibnamefont {Ogawa}}, \bibinfo {author} {\bibfnamefont {A.}~\bibnamefont {Kadokura}}, \bibinfo {author} {\bibfnamefont {H.}~\bibnamefont {Miyaoka}},\ and\ \bibinfo {author} {\bibfnamefont {N.}~\bibnamefont {Sato}},\ }\bibfield  {title} {\bibinfo {title} {Modulation of ionospheric conductance and electric field associated with pulsating aurora},\ }\bibfield  {journal} {\bibinfo  {journal} {Journal of Geophysical Research: Space Physics}\ }\textbf {\bibinfo {volume} {115}},\ \href {https://doi.org/10.1029/2009JA014683} {10.1029/2009JA014683} (\bibinfo {year} {2010})\BibitemShut {NoStop}%
\bibitem [{\citenamefont {Hosokawa}\ \emph {et~al.}(2013)\citenamefont {Hosokawa}, \citenamefont {Milan}, \citenamefont {Lester}, \citenamefont {Kadokura}, \citenamefont {Sato},\ and\ \citenamefont {Bjornsson}}]{hosokawaLargeFlowShears2013}%
  \BibitemOpen
  \bibfield  {author} {\bibinfo {author} {\bibfnamefont {K.}~\bibnamefont {Hosokawa}}, \bibinfo {author} {\bibfnamefont {S.~E.}\ \bibnamefont {Milan}}, \bibinfo {author} {\bibfnamefont {M.}~\bibnamefont {Lester}}, \bibinfo {author} {\bibfnamefont {A.}~\bibnamefont {Kadokura}}, \bibinfo {author} {\bibfnamefont {N.}~\bibnamefont {Sato}},\ and\ \bibinfo {author} {\bibfnamefont {G.}~\bibnamefont {Bjornsson}},\ }\bibfield  {title} {\bibinfo {title} {Large flow shears around auroral beads at substorm onset},\ }\href@noop {} {\bibfield  {journal} {\bibinfo  {journal} {Geophysical Research Letters}\ }\textbf {\bibinfo {volume} {40}},\ \bibinfo {pages} {4987} (\bibinfo {year} {2013})}\BibitemShut {NoStop}%
\bibitem [{\citenamefont {Ivarsen}\ \emph {et~al.}(2025{\natexlab{b}})\citenamefont {Ivarsen}, \citenamefont {Miyashita}, \citenamefont {{St-Maurice}}, \citenamefont {Hussey}, \citenamefont {Pitzel}, \citenamefont {Galeschuk}, \citenamefont {Marei}, \citenamefont {Horne}, \citenamefont {Kasahara}, \citenamefont {Matsuda}, \citenamefont {Kasahara}, \citenamefont {Keika}, \citenamefont {Miyoshi}, \citenamefont {Yamamoto}, \citenamefont {Shinbori}, \citenamefont {Huyghebaert}, \citenamefont {Matsuoka}, \citenamefont {Yokota},\ and\ \citenamefont {Tsuchiya}}]{ivarsen_characteristic_2025}%
  \BibitemOpen
  \bibfield  {author} {\bibinfo {author} {\bibfnamefont {M.~F.}\ \bibnamefont {Ivarsen}}, \bibinfo {author} {\bibfnamefont {Y.}~\bibnamefont {Miyashita}}, \bibinfo {author} {\bibfnamefont {J.-P.}\ \bibnamefont {{St-Maurice}}}, \bibinfo {author} {\bibfnamefont {G.~C.}\ \bibnamefont {Hussey}}, \bibinfo {author} {\bibfnamefont {B.}~\bibnamefont {Pitzel}}, \bibinfo {author} {\bibfnamefont {D.}~\bibnamefont {Galeschuk}}, \bibinfo {author} {\bibfnamefont {S.}~\bibnamefont {Marei}}, \bibinfo {author} {\bibfnamefont {R.~B.}\ \bibnamefont {Horne}}, \bibinfo {author} {\bibfnamefont {Y.}~\bibnamefont {Kasahara}}, \bibinfo {author} {\bibfnamefont {S.}~\bibnamefont {Matsuda}}, \bibinfo {author} {\bibfnamefont {S.}~\bibnamefont {Kasahara}}, \bibinfo {author} {\bibfnamefont {K.}~\bibnamefont {Keika}}, \bibinfo {author} {\bibfnamefont {Y.}~\bibnamefont {Miyoshi}}, \bibinfo {author} {\bibfnamefont {K.}~\bibnamefont {Yamamoto}}, \bibinfo {author} {\bibfnamefont {A.}~\bibnamefont {Shinbori}}, \bibinfo {author} {\bibfnamefont
  {D.~R.}\ \bibnamefont {Huyghebaert}}, \bibinfo {author} {\bibfnamefont {A.}~\bibnamefont {Matsuoka}}, \bibinfo {author} {\bibfnamefont {S.}~\bibnamefont {Yokota}},\ and\ \bibinfo {author} {\bibfnamefont {F.}~\bibnamefont {Tsuchiya}},\ }\bibfield  {title} {\bibinfo {title} {Characteristic {{E-Region Plasma Signature}} of {{Magnetospheric Wave-Particle Interactions}}},\ }\href {https://doi.org/10.1103/PhysRevLett.134.145201} {\bibfield  {journal} {\bibinfo  {journal} {Physical Review Letters}\ }\textbf {\bibinfo {volume} {134}},\ \bibinfo {pages} {145201} (\bibinfo {year} {2025}{\natexlab{b}})}\BibitemShut {NoStop}%
\bibitem [{\citenamefont {Feng}\ \emph {et~al.}(2024)\citenamefont {Feng}, \citenamefont {Wang}, \citenamefont {Guo}, \citenamefont {Shprits}, \citenamefont {Han}, \citenamefont {Teng}, \citenamefont {Ni}, \citenamefont {Shi},\ and\ \citenamefont {Zhang}}]{feng_lower_2024}%
  \BibitemOpen
  \bibfield  {author} {\bibinfo {author} {\bibfnamefont {H.}~\bibnamefont {Feng}}, \bibinfo {author} {\bibfnamefont {D.}~\bibnamefont {Wang}}, \bibinfo {author} {\bibfnamefont {D.}~\bibnamefont {Guo}}, \bibinfo {author} {\bibfnamefont {Y.~Y.}\ \bibnamefont {Shprits}}, \bibinfo {author} {\bibfnamefont {D.}~\bibnamefont {Han}}, \bibinfo {author} {\bibfnamefont {S.}~\bibnamefont {Teng}}, \bibinfo {author} {\bibfnamefont {B.}~\bibnamefont {Ni}}, \bibinfo {author} {\bibfnamefont {R.}~\bibnamefont {Shi}},\ and\ \bibinfo {author} {\bibfnamefont {Y.}~\bibnamefont {Zhang}},\ }\bibfield  {title} {\bibinfo {title} {Lower {{Band Chorus Wave Scattering Causing}} the {{Extensive Morningside Diffuse Auroral Precipitation During Active Geomagnetic Conditions}}: {{A Detailed Case Study}}},\ }\href {https://doi.org/10.1029/2023JA032240} {\bibfield  {journal} {\bibinfo  {journal} {Journal of Geophysical Research: Space Physics}\ }\textbf {\bibinfo {volume} {129}},\ \bibinfo {pages} {e2023JA032240} (\bibinfo {year}
  {2024})}\BibitemShut {NoStop}%
\bibitem [{\citenamefont {Palmroth}\ \emph {et~al.}(2021)\citenamefont {Palmroth}, \citenamefont {Grandin}, \citenamefont {Sarris}, \citenamefont {Doornbos}, \citenamefont {Tourgaidis}, \citenamefont {Aikio}, \citenamefont {Buchert}, \citenamefont {Clilverd}, \citenamefont {Dandouras}, \citenamefont {Heelis}, \citenamefont {Hoffmann}, \citenamefont {Ivchenko}, \citenamefont {Kervalishvili}, \citenamefont {Knudsen}, \citenamefont {Kotova}, \citenamefont {Liu}, \citenamefont {Malaspina}, \citenamefont {March}, \citenamefont {Marchaudon}, \citenamefont {Marghitu}, \citenamefont {Matsuo}, \citenamefont {Miloch}, \citenamefont {{Moretto-J{\o}rgensen}}, \citenamefont {Mpaloukidis}, \citenamefont {Olsen}, \citenamefont {Papadakis}, \citenamefont {Pfaff}, \citenamefont {Pirnaris}, \citenamefont {Siemes}, \citenamefont {Stolle}, \citenamefont {Suni}, \citenamefont {{van den IJssel}}, \citenamefont {Verronen}, \citenamefont {Visser},\ and\ \citenamefont {Yamauchi}}]{palmrothLowerthermosphereIonosphereLTI2021}%
  \BibitemOpen
  \bibfield  {author} {\bibinfo {author} {\bibfnamefont {M.}~\bibnamefont {Palmroth}}, \bibinfo {author} {\bibfnamefont {M.}~\bibnamefont {Grandin}}, \bibinfo {author} {\bibfnamefont {T.}~\bibnamefont {Sarris}}, \bibinfo {author} {\bibfnamefont {E.}~\bibnamefont {Doornbos}}, \bibinfo {author} {\bibfnamefont {S.}~\bibnamefont {Tourgaidis}}, \bibinfo {author} {\bibfnamefont {A.}~\bibnamefont {Aikio}}, \bibinfo {author} {\bibfnamefont {S.}~\bibnamefont {Buchert}}, \bibinfo {author} {\bibfnamefont {M.~A.}\ \bibnamefont {Clilverd}}, \bibinfo {author} {\bibfnamefont {I.}~\bibnamefont {Dandouras}}, \bibinfo {author} {\bibfnamefont {R.}~\bibnamefont {Heelis}}, \bibinfo {author} {\bibfnamefont {A.}~\bibnamefont {Hoffmann}}, \bibinfo {author} {\bibfnamefont {N.}~\bibnamefont {Ivchenko}}, \bibinfo {author} {\bibfnamefont {G.}~\bibnamefont {Kervalishvili}}, \bibinfo {author} {\bibfnamefont {D.~J.}\ \bibnamefont {Knudsen}}, \bibinfo {author} {\bibfnamefont {A.}~\bibnamefont {Kotova}}, \bibinfo {author} {\bibfnamefont
  {H.-L.}\ \bibnamefont {Liu}}, \bibinfo {author} {\bibfnamefont {D.~M.}\ \bibnamefont {Malaspina}}, \bibinfo {author} {\bibfnamefont {G.}~\bibnamefont {March}}, \bibinfo {author} {\bibfnamefont {A.}~\bibnamefont {Marchaudon}}, \bibinfo {author} {\bibfnamefont {O.}~\bibnamefont {Marghitu}}, \bibinfo {author} {\bibfnamefont {T.}~\bibnamefont {Matsuo}}, \bibinfo {author} {\bibfnamefont {W.~J.}\ \bibnamefont {Miloch}}, \bibinfo {author} {\bibfnamefont {T.}~\bibnamefont {{Moretto-J{\o}rgensen}}}, \bibinfo {author} {\bibfnamefont {D.}~\bibnamefont {Mpaloukidis}}, \bibinfo {author} {\bibfnamefont {N.}~\bibnamefont {Olsen}}, \bibinfo {author} {\bibfnamefont {K.}~\bibnamefont {Papadakis}}, \bibinfo {author} {\bibfnamefont {R.}~\bibnamefont {Pfaff}}, \bibinfo {author} {\bibfnamefont {P.}~\bibnamefont {Pirnaris}}, \bibinfo {author} {\bibfnamefont {C.}~\bibnamefont {Siemes}}, \bibinfo {author} {\bibfnamefont {C.}~\bibnamefont {Stolle}}, \bibinfo {author} {\bibfnamefont {J.}~\bibnamefont {Suni}}, \bibinfo {author}
  {\bibfnamefont {J.}~\bibnamefont {{van den IJssel}}}, \bibinfo {author} {\bibfnamefont {P.~T.}\ \bibnamefont {Verronen}}, \bibinfo {author} {\bibfnamefont {P.}~\bibnamefont {Visser}},\ and\ \bibinfo {author} {\bibfnamefont {M.}~\bibnamefont {Yamauchi}},\ }\bibfield  {title} {\bibinfo {title} {Lower-thermosphere--ionosphere ({{LTI}}) quantities: Current status of measuring techniques and models},\ }\href {https://doi.org/10.5194/angeo-39-189-2021} {\bibfield  {journal} {\bibinfo  {journal} {Annales Geophysicae}\ }\textbf {\bibinfo {volume} {39}},\ \bibinfo {pages} {189} (\bibinfo {year} {2021})}\BibitemShut {NoStop}%
\bibitem [{\citenamefont {Wu}\ \emph {et~al.}(2022)\citenamefont {Wu}, \citenamefont {Wang}, \citenamefont {Lin}, \citenamefont {Huang},\ and\ \citenamefont {Zhang}}]{wu_penetrating_2022}%
  \BibitemOpen
  \bibfield  {author} {\bibinfo {author} {\bibfnamefont {Q.}~\bibnamefont {Wu}}, \bibinfo {author} {\bibfnamefont {W.}~\bibnamefont {Wang}}, \bibinfo {author} {\bibfnamefont {D.}~\bibnamefont {Lin}}, \bibinfo {author} {\bibfnamefont {C.}~\bibnamefont {Huang}},\ and\ \bibinfo {author} {\bibfnamefont {Y.}~\bibnamefont {Zhang}},\ }\bibfield  {title} {\bibinfo {title} {Penetrating {{Electric Field Simulated}} by the {{MAGE}} and {{Comparison With ICON Observation}}},\ }\href {https://doi.org/10.1029/2022JA030467} {\bibfield  {journal} {\bibinfo  {journal} {Journal of Geophysical Research: Space Physics}\ }\textbf {\bibinfo {volume} {127}},\ \bibinfo {pages} {e2022JA030467} (\bibinfo {year} {2022})}\BibitemShut {NoStop}%
\bibitem [{\citenamefont {Wiltberger}\ \emph {et~al.}(2017)\citenamefont {Wiltberger}, \citenamefont {Merkin}, \citenamefont {Zhang}, \citenamefont {Toffoletto}, \citenamefont {Oppenheim}, \citenamefont {Wang}, \citenamefont {Lyon}, \citenamefont {Liu}, \citenamefont {Dimant}, \citenamefont {Sitnov},\ and\ \citenamefont {Stephens}}]{wiltbergerEffectsElectrojetTurbulence2017}%
  \BibitemOpen
  \bibfield  {author} {\bibinfo {author} {\bibfnamefont {M.}~\bibnamefont {Wiltberger}}, \bibinfo {author} {\bibfnamefont {V.}~\bibnamefont {Merkin}}, \bibinfo {author} {\bibfnamefont {B.}~\bibnamefont {Zhang}}, \bibinfo {author} {\bibfnamefont {F.}~\bibnamefont {Toffoletto}}, \bibinfo {author} {\bibfnamefont {M.}~\bibnamefont {Oppenheim}}, \bibinfo {author} {\bibfnamefont {W.}~\bibnamefont {Wang}}, \bibinfo {author} {\bibfnamefont {J.~G.}\ \bibnamefont {Lyon}}, \bibinfo {author} {\bibfnamefont {J.}~\bibnamefont {Liu}}, \bibinfo {author} {\bibfnamefont {Y.}~\bibnamefont {Dimant}}, \bibinfo {author} {\bibfnamefont {M.~I.}\ \bibnamefont {Sitnov}},\ and\ \bibinfo {author} {\bibfnamefont {G.~K.}\ \bibnamefont {Stephens}},\ }\bibfield  {title} {\bibinfo {title} {Effects of electrojet turbulence on a magnetosphere-ionosphere simulation of a geomagnetic storm},\ }\href {https://doi.org/10.1002/2016JA023700} {\bibfield  {journal} {\bibinfo  {journal} {Journal of Geophysical Research: Space Physics}\ }\textbf
  {\bibinfo {volume} {122}},\ \bibinfo {pages} {5008} (\bibinfo {year} {2017})}\BibitemShut {NoStop}%
\bibitem [{\citenamefont {{St-Maurice}}\ and\ \citenamefont {Goodwin}(2021)}]{st-mauriceRevisitingBehaviorERegion2021}%
  \BibitemOpen
  \bibfield  {author} {\bibinfo {author} {\bibfnamefont {J.-P.}\ \bibnamefont {{St-Maurice}}}\ and\ \bibinfo {author} {\bibfnamefont {L.}~\bibnamefont {Goodwin}},\ }\bibfield  {title} {\bibinfo {title} {Revisiting the {{Behavior}} of the {{E-Region Electron Temperature During Strong Electric Field Events}} at {{High Latitudes}}},\ }\href {https://doi.org/10.1029/2020JA028288} {\bibfield  {journal} {\bibinfo  {journal} {Journal of Geophysical Research: Space Physics}\ }\textbf {\bibinfo {volume} {126}},\ \bibinfo {pages} {2020JA028288} (\bibinfo {year} {2021})}\BibitemShut {NoStop}%
\bibitem [{\citenamefont {Redmon}\ \emph {et~al.}(2017)\citenamefont {Redmon}, \citenamefont {Denig}, \citenamefont {Kilcommons},\ and\ \citenamefont {Knipp}}]{redmonNewDMSPDatabase2017}%
  \BibitemOpen
  \bibfield  {author} {\bibinfo {author} {\bibfnamefont {R.~J.}\ \bibnamefont {Redmon}}, \bibinfo {author} {\bibfnamefont {W.~F.}\ \bibnamefont {Denig}}, \bibinfo {author} {\bibfnamefont {L.~M.}\ \bibnamefont {Kilcommons}},\ and\ \bibinfo {author} {\bibfnamefont {D.~J.}\ \bibnamefont {Knipp}},\ }\bibfield  {title} {\bibinfo {title} {New {{DMSP}} database of precipitating auroral electrons and ions},\ }\href {https://doi.org/10.1002/2016JA023339} {\bibfield  {journal} {\bibinfo  {journal} {Journal of Geophysical Research: Space Physics}\ }\textbf {\bibinfo {volume} {122}},\ \bibinfo {pages} {9056} (\bibinfo {year} {2017})}\BibitemShut {NoStop}%
\bibitem [{\citenamefont {Newell}\ \emph {et~al.}(2010)\citenamefont {Newell}, \citenamefont {Sotirelis},\ and\ \citenamefont {Wing}}]{newellSeasonalVariationsDiffuse2010}%
  \BibitemOpen
  \bibfield  {author} {\bibinfo {author} {\bibfnamefont {P.~T.}\ \bibnamefont {Newell}}, \bibinfo {author} {\bibfnamefont {T.}~\bibnamefont {Sotirelis}},\ and\ \bibinfo {author} {\bibfnamefont {S.}~\bibnamefont {Wing}},\ }\bibfield  {title} {\bibinfo {title} {Seasonal variations in diffuse, monoenergetic, and broadband aurora},\ }\bibfield  {journal} {\bibinfo  {journal} {Journal of Geophysical Research: Space Physics}\ }\textbf {\bibinfo {volume} {115}},\ \href {https://doi.org/10.1029/2009JA014805} {10.1029/2009JA014805} (\bibinfo {year} {2010})\BibitemShut {NoStop}%
\bibitem [{\citenamefont {Ivarsen}\ \emph {et~al.}(2024{\natexlab{c}})\citenamefont {Ivarsen}, \citenamefont {{St-Maurice}}, \citenamefont {Jin}, \citenamefont {Park}, \citenamefont {Buschman},\ and\ \citenamefont {Clausen}}]{ivarsen_what_2024}%
  \BibitemOpen
  \bibfield  {author} {\bibinfo {author} {\bibfnamefont {M.~F.}\ \bibnamefont {Ivarsen}}, \bibinfo {author} {\bibfnamefont {J.-P.}\ \bibnamefont {{St-Maurice}}}, \bibinfo {author} {\bibfnamefont {Y.}~\bibnamefont {Jin}}, \bibinfo {author} {\bibfnamefont {J.}~\bibnamefont {Park}}, \bibinfo {author} {\bibfnamefont {L.~M.}\ \bibnamefont {Buschman}},\ and\ \bibinfo {author} {\bibfnamefont {L.~B.}\ \bibnamefont {Clausen}},\ }\bibfield  {title} {\bibinfo {title} {To what degree does the high-energy aurora destroy {{F-region}} irregularities?},\ }\bibfield  {journal} {\bibinfo  {journal} {Frontiers in Astronomy and Space Sciences}\ }\textbf {\bibinfo {volume} {11}},\ \href {https://doi.org/10.3389/fspas.2024.1309136} {10.3389/fspas.2024.1309136} (\bibinfo {year} {2024}{\natexlab{c}})\BibitemShut {NoStop}%
\end{thebibliography}

%

\end{document}